\documentclass[a4paper,12pt]{article}

\usepackage{graphics, epsfig}

\begin{document}

\author{E. M. O'Shea \thanks{E-mail : emer.oshea@ucd.ie}\\
\small Mathematical Physics Department\\ \small  National
University of Ireland Dublin, Belfield, Dublin 4, Ireland}

\title{Metric Perturbation Approach to Gravitational Waves in Isotropic Cosmologies}
\date{}
\maketitle

\begin{abstract}
Gravitational waves in isotropic cosmologies were recently studied
using the gauge--invariant approach of Ellis--Bruni
\cite{isotropic}. We now construct the linearised metric
perturbations of the background Robertson--Walker space--time
which reproduce the results obtained in that study. The analysis
carried out here also facilitates an easy comparison with Bardeen.
\end{abstract}
\vskip 2truepc\noindent PACS number(s): 04.30.Nk \vskip
6truepc\noindent Accepted for publication in Phys. Rev. D
\thispagestyle{empty}
\newpage

\section{Introduction}\indent
In a recent paper \cite{isotropic} the gauge--invariant and
covariant approach of Ellis--Bruni \cite{EB} is used to examine
shear--free gravitational waves propagating through isotropic
cosmologies. In this approach the waves are modelled as small
perturbations of the Robertson--Walker space--time. The presence
of the waves is found to perturb the shear and also more notably
to introduce anisotropic stress into the universe. Other basic
gauge--invariant quantities, for example the vorticity and energy
flow, remain unchanged by the presence of gravitational radiation.

Our purpose here is to construct the metric perturbations of the
Robertson--Walker space--time which give rise to the perturbations
of the anisotropic stress and shear found in \cite{isotropic}. The
difficulty is that we wish to derive gauge--invariant
perturbations and there is no way a priori to identify which terms
in the perturbed metric are pure gauge terms without carrying out
a lengthy calculation. In the process of studying the perturbed
metric we identify the gauge terms and without loss of generality
we then put these terms equal to zero.

The paper is organised as follows: In Section 2 we introduce the
notation used and give some important equations. The unperturbed
Robertson--Walker space--time  is described in Section 3. In
Section 4 we summarise the results of the gauge--invariant and
covariant study of gravitational radiation carried out in
\cite{isotropic}. The perturbed metric is introduced in Section 5.
Also in this section and Section 6 we demonstrate how the
perturbed metric leads to the required gauge--invariant
perturbations of the shear and anisotropic stress. The Ricci
tensor components of the metric are listed in Appendix A and in
Appendix B we briefly outline the calculation involved in
identifying those variables which are responsible for the presence
of gauge terms. The paper ends with a discussion in which our
results are compared with those of Bardeen \cite{bardeen}.

\setcounter{equation}{0}
\section{Notation and Basic Equations}\indent

Throughout this paper we use the notation and sign conventions of
\cite{Ellis}. We are concerned with a four dimensional space--time
manifold with metric tensor components $g_{ab}$ in a local
coordinate system $\{x^a\}$ and a preferred congruence of
world--lines tangent to a time--like vector field with components
$u^a$ and $u^au_a=-1$. With respect to this 4--velocity field the
symmetric energy-momentum-stress tensor $T^{ab}$ can be decomposed
as
\begin{equation}\label {2.1}
T^{ab} = \mu\,u^a\,u^b+p\,h^{ab}+q^a\,u^b+q^b\,u^a+\pi^{ab}\ ,
\end{equation}
where
\begin{equation}\label{2.2}
h^{ab}=g^{ab}+u^a\,u^b\ ,
\end{equation}
is the projection tensor and
\begin{equation}\label{2.3}
q^a\,u_a=0\ , \qquad \pi^{ab}\,u_a=0\ , \qquad {\pi^a}_a=0\ ,
\end{equation}
with $\pi^{ab}=\pi^{ba}$. Here $\mu$ is the matter energy density
measured by the observer with 4--velocity $u^a$, $p$ is the
isotropic pressure, $q^a$ is the energy flow relative to $u^a$
(for example heat flow) and $\pi^{ab}$ is the trace--free
anisotropic stress (due to processes such as viscosity).

We indicate covariant differentiation with a semicolon, partial
differentiation by a comma and covariant differentiation in the
direction of $u^a$ by a dot. Also as usual square brackets denote
skew--symmetrization, round brackets denote symmetrization and a
definition is indicated by a colon followed by an equality sign.
Thus the 4--acceleration of the time--like congruence is
\begin{equation}\label{2.4}
\dot{u}^a={u^a}_{;\,b}\,u^b\ ,
\end{equation}
and $u_{a\,;\,b}$ can be decomposed into
\begin{equation}\label{2.5}
u_{a\,;\,b}=\omega_{ab}+\sigma_{ab}+\frac{1}{3}\,\theta\,h_{ab}-{\dot{u}}_a\,u_b\
,
\end{equation}
where
\begin{equation}\label{2.6}
\omega_{ab}:= u_{[\,a\,;\,b\,]}+{\dot{u}}_{[\,a}\,u_{b\,]}\ ,
\end{equation}
is the vorticity tensor of the congruence,
\begin{equation}\label{2.7}
\sigma_{ab}:=u_{(\,a\,;\,b\,)}+{\dot{u}}_{(\,a}\,u_{b\,)}-\frac{1}{3}\,\theta\,h_{ab}\
,
\end{equation}
is the shear tensor of the congruence and
\begin{equation}\label{2.8}
\theta:={u^a}_{;\,a}\ ,
\end{equation}
is the expansion (or contraction) of the congruence.

We shall make use of the Ricci identities
\begin{equation}\label{2.9}
u_{a\,;\,d\,c}-u_{a\,;\,c\,d}=R_{abcd}\,u^b\ ,
\end{equation}
where $R_{abcd}$ is the Riemann curvature tensor but for the
problem at hand the key equations are Einstein's field equations
\begin{equation}\label{2.10}
R_{ab}-\frac{1}{2}\,g_{ab}\,R=T_{ab}\ .
\end{equation}
Here $R_{ab}:={{R_a}^c}_{bc}$ are the components of the Ricci
tensor, $R:={R^c}_c$ is the Ricci scalar and we have absorbed the
coupling constant into the energy--momentum--stress tensor. Noting
that $R=-\,T\,(:={T^a}_a)$ and using Eq. (\ref{2.1}) the field
equations can  be decomposed into:
\begin{eqnarray}\label{2.11}
R_{ab}\,u^a\,u^b &=& \frac{1}{2}\,(\mu+3\,p)\ ,\nonumber \\
R_{ab}\,u^a\,h^b_c&=&\,-\,q_c\ ,  \\
R_{ab}\,h^a_c\,h^b_d&=&\,\frac{1}{2}\,(\mu-p)\,h_{cd}+\pi_{cd}\
.\nonumber
\end{eqnarray}
It is in this form that we shall use Eq. (\ref{2.10}) in later
sections.

 \setcounter{equation}{0}
\section{The Background Space--Time}\indent

We choose as the unperturbed (background) space--time a
Robertson--Walker space--time with line--element
\begin{equation}\label{3.1}
ds^2=R^2(t)\,\frac{[(dx^1)^2+(dx^2)^2+
(dx^3)^2]}{\left(1+\frac{k}{4}\,r^2\right)^2}-dt^2\ ,
\end{equation}
where $R(t)$ is the scale factor, $r^2=(x^1)^2+(x^2)^2+(x^3)^2$
and $k=0,\pm 1$ is the Gaussian curvature of the space--like
hypersurfaces $t= \rm {const}$. The world--lines of the fluid
particles are the integral curves of the vector field
$u^a\,\partial/\partial x^a\,=\,\partial/\partial t$ (thus
$u^a=\delta^a_4 $ since we shall label the coordinates $ x^1=y\ ,
x^2=z\ , x^3=x\ ,x^4=t$). The background energy--momentum--stress
tensor is Eq. (\ref{2.1}) specialized to a perfect fluid (by
putting $q^a=0=\pi^{ab}$) with proper--density
\begin{equation}\label{3.2}
\mu=3\,\frac{\dot{R}^2}{R^2}+3\,\frac{k}{R^2}\ ,
\end{equation}
and isotropic pressure
\begin{equation}\label{3.3}
p=-\frac{\dot{R}^2}{R^2}-2\,\frac{\ddot{R}}{R}-\frac{k}{R^2}\ .
\end{equation}

We find it convenient to put the line--element given above in the
following forms:
\begin{equation}\label{3.4}
ds^2=R^2(t)\,\{dx^2+p_0^{-2}\,f^2\,(dy^2+dz^2)\}-dt^2\ ,
\end{equation}
with $p_0=1+(K/4)(y^2+z^2)$, $K=\rm const$, $f=f(x)$. We identify
three distinct cases:\\ \textbf{Case 1:} \\ If $k=+1$ then $K=+1$
and $f(x)=\sin{x}$.\\ Noting that the transformation
$x\rightarrow\pi/2-x$ does not affect the \emph{form} of the
line--element (\ref{3.4}) we see that in this case $f(x)$ could
equivalently be written $f(x)=\cos{x}$
\\ \\
\textbf{Case 2:} \\
$\mbox{If }k=0 \mbox{ then }\left\{ \begin{array}{lcl} K=0 &
\rm{and} &
f(x)=1\ , \\
{}& \rm{or} & {} \\
K=+1 & \rm{and} & f(x)=x\ .
\end{array} \right.  $
\\ \\ \\
\textbf{Case 3:} \\
$\mbox{If }k=-1 \mbox{ then }\left\{ \begin{array}{lcl} K=-1 &
\rm{and} &
f(x)=\cosh x\ ,\\
{}& \rm{or} & {} \\
K=0 & \rm{and} & f(x)=\frac{1}{2}\,e^x \ ,\\
{}& \rm{or} & {} \\
K=+1 & \rm{and} & f(x)=\sinh x\ .
\end{array} \right.  $
\\
The form of the line--element (\ref{3.4}) is also invariant under
the transformation $x\rightarrow\,-x$ so when $K=0$ in case 3 we
could instead write $f(x)=\frac{1}{2}\,e^{-x}$. For a detailed
explanation why these cases arise see for example Eqs.
(5.3)--(5.19) in \cite{isotropic}. In space--times with
line--elements (\ref{3.4}) the hypersurfaces
\begin{equation}\label{3.5}
\phi(x^a) := x-T(t) = \rm {const}\ ,
\end{equation}
with $dT/dt = R^{-1}$ are null hypersurfaces. The expansion of the
null geodesic generators of these surfaces is
\begin{equation}\label{3.6}
\frac{1}{2}\,\phi ^{,a}{}_{;a}=\frac{f'}{R^2f}+\frac{\dot R}{R^2}\
,
\end{equation}
where $f^{'} = df/dx$, $\dot R = dR/dt$. Using (\ref{3.5}) we can
show that
\begin{equation}\label{3.7}
2\phi_{,a;b}=\xi_a\,\phi_{,b}+\xi_b\,\phi_{,a}+\phi_{,d}{}^{;d}\,g_{ab}\
,
\end{equation}
where \begin{equation}\label{3.8} \xi_{a}=
-\,\frac{f^{'}}{f}\,\phi_{,a}+R\,\phi_{,d}{}^{;d}\,u_{a}\ .
\end{equation}
It follows from Eq. (\ref{3.7}) that $\phi_{,a}$ is shear--free
\cite{robinson}.

Finally in this section we note that for convenience we have used
the same coordinate labels $\{y, z, x, t\}$ for all the special
cases included in (\ref{3.4}). Clearly the ranges of some of these
coordinates will vary from case to case and within cases 2 and 3.
For example, in case 2 $x\in\,(-\infty\, , +\infty)$ if $K=0$ but
$x\in[\,0,+\infty)$ and is a radial polar coordinate if $K=+1$.
The shear--free null hypersurfaces (\ref{3.5}) will also be
different in the different cases. This can be seen by examining
the intersections of these null hypersurfaces with the space--like
hypersurfaces $t=\rm {const}$.\newline \textbf{Case 1}\newline The
intersection is a 2--sphere.\newline\newline \textbf{Case
2}\newline If $K=+1$ the intersection is a 2--sphere and if $K=0$
the intersection is a 2--plane. Thus it is obvious that Eq.
(\ref{3.5}) describes two different families of shear--free null
hypersurfaces that can occur in an open, spatially flat
universe.\newline\newline \textbf{Case 3}\newline In this case the
intersection of (\ref{3.5}) with the $t=\rm {const}$ hypersurfaces
is always a 2--space of constant curvature. The curvature of this
2--space is given by $K$ which takes values $0,\pm 1$. So we have
three different families of shear--free null hypersurfaces in a
$k=-1$ universe. We refer the reader to \cite{Hogan} for a
geometrical explanation for the existence of these subcases.

\setcounter{equation}{0}
\section{Gauge--Invariant and Covariant Approach to Gravitational Waves}\indent

In a recent paper \cite{isotropic} we used the gauge--invariant
and covariant approach of Ellis--Bruni \cite{EB} to construct
gravitational wave perturbations of the Robertson--Walker
space--times described in the previous section. This involves
working in a general local coordinate system with gauge--invariant
small quantities which by their nature vanish in the background,
rather than small perturbations of the background metric. For
isotropic space--times the Ellis--Bruni variables are
$\sigma_{ab}$, $\dot{u}^a$, $\omega_{ab}$,
$X_{a}=h^b_a\,\mu_{,b}$, $Y_a=h^b_a\,p_{,b}$,
$Z_a=h^b_a\,\theta_{,b}$, $\pi_{ab}$, $q_a$ and the ``electric"
and ``magnetic" parts of the Weyl tensor, with components
$C_{abcd}$, given respectively by
\begin{equation}\label{4.1}
E_{ab}=C_{apbq}\,u^p\,u^q\ , \qquad
H_{ab}={}^*\,C_{apbq}\,u^p\,u^q\ .
\end{equation}
Here ${}^*C_{apbq}=\frac{1}{2}\,\eta_{ap}{}^{rs}\,C_{rsbq}$ is the
dual of the Weyl tensor (the left and right duals being equal),
$\eta_{abcd}=\sqrt{-g}\,\epsilon_{abcd}$ where
$g=\rm{det}(g_{ab})$ and $\epsilon_{abcd}$ is the Levi--Civita
permutation symbol. However we found that it is tensor quantities
that describe gravitational wave perturbations. Thus for this
problem the important Ellis--Bruni variables are $\sigma_{ab}$,
$\pi_{ab}$, $E_{ab}$, $H_{ab}$ and we can set all other
gauge--invariant variables equal to zero.  The equations satisfied
by these variables are obtained by projections in the direction
$u^a$ and orthogonal to $u^a$ of the Ricci identities, the
equations of motion and the energy conservation equation contained
in $T^{ab}{}_{;b}=0$ and the Bianchi identities written in the
form
\begin{equation}\label{4.2}
C^{abcd}{}_{;d}=R^{c[a;b]}-\frac{1}{6}\,g^{c[a}\,R^{;b]}\ .
\end{equation}
To keep this section to a reasonable length we shall not list all
of the equations (they are given in Eqs. (2.14)--(2.25) in
\cite{isotropic}). We note here that from the projections of the
Ricci identities (after putting $\dot{u}^a=0=\omega_{ab}$) we find
\begin{equation}\label{4.3}
E_{ab}=\frac{1}{2}\,\pi_{ab}+\frac{2}{3}\,\sigma^2\,h_{ab}-\frac{2}{3}\,\theta\,\sigma_{ab}-\sigma_{af}\,\sigma^f{}_b-
h^f_a\,h^g_b\,\dot{\sigma}_{ab}\ ,
\end{equation}
and
\begin{equation}\label{4.4}
H_{ab}=-\,h^t_a\,h^s_b\,\sigma_{(t}{}^{g;c}\,\eta_{s)fgc}\,u^f\ .
\end{equation}
Thus these variables are derived from $\pi_{ab}$ and
$\sigma_{ab}$.

We now assume that the perturbed shear and anisotropic stress have
the following form:
\begin{equation}\label{4.5}
\sigma_{ab}=s_{ab}\,F(\phi)\ , \qquad \pi_{ab}=\Pi_{ab}\,F(\phi)\
,
\end{equation}
where $F$ is an arbitrary real--valued function of its argument
$\phi(x^a)$. We emphasise that at this point $\phi(x^a)$ is
arbitrary and not that defined in Eq. (\ref{3.5}). This idea of
introducing arbitrary functions into solutions of Einstein's
equations describing gravitational waves goes back to work by
Trautman \cite{Traut} and the above form for the gauge--invariant
variables was introduced by Hogan and Ellis \cite{H+E}.
Substituting (\ref{4.5}) into the linearised versions of the
equations satisfied by these variables and noting that $s_{ab}$
and $\Pi_{ab}$ are trace--free and orthogonal to $u_a$ with
respect to the background metric we find that \cite{isotropic}
\begin{equation}\label{4.6}
g^{ab}\,\phi _{,a}\,\phi _{,b}=0\ ,\qquad s^{ab}\,\phi_{,b}=0\ ,
\qquad \Pi^{ab}\,\phi_{,b}=0\ ,
\end{equation}
with $g_{ab}$ here the background metric, and
\begin{equation}\label{4.7}
s^{ab}{}_{|b}=0\ , \qquad \Pi^{ab}{}_{|b}=0\ ,
\end{equation}
where for clarity we have used a stroke to denote covariant
differentiation with respect to the background metric. We also
discover (see \cite{isotropic}) the following wave equation for
$s_{ab}$
\begin{equation}\label{4.8}
s^{ab|d}{}_{|d}-\frac{2}{3}\,\theta\,\dot s^{ab}-\left
(\frac{1}{3}\,\dot\theta +\frac{4}{9}\,\theta ^2\right
)\,s^{ab}+(p-\frac{1}{3}\,\mu )\, s^{ab}=-\dot\Pi
^{ab}-\frac{2}{3}\,\theta\,\Pi ^{ab}\ ,
\end{equation}
and a propagation equation for $s_{ab}$ along the null geodesics
tangent to $\phi^{,d}$, namely,
\begin{equation} \label{4.9}
s'_{tb}+\left (\frac{1}{2}\,\phi
^{,d}{}_{|d}-\frac{1}{3}\,\theta\,\dot\phi \right
)\,s_{tb}=-\frac{1}{2}\,\dot\phi\,\Pi _{tb}\ ,
\end{equation}
where $s^{'}_{tb}:=s_{tb|d}\,\phi^{,d}$ and
$\dot{\phi}=\phi_{,a}\,u^a$. The internal consistencies of these
equations were checked in \cite{isotropic}. The ``electric" and
``magnetic" parts of the Weyl tensor are now given by
\cite{isotropic}
\begin{equation}\label{4.10}
E_{ab}=\left(\frac{1}{2}\,\Pi_{ab}-\dot{s}_{ab}-\frac{2}{3}\,\theta\,s_{ab}\,\right)\,F-\dot{\phi}\,s_{ab}\,F^{'}
\
,
\end{equation}
and
\begin{equation}\label {4.11}
H_{ab}=-s_{(a}{}^{p|c}\,\eta _{b)fpc}\,u^f\,F-s_{(a}{}^p\,\eta
_{b)fpc}\,u^f\,\phi ^{,c}\,F^{'}\ ,
\end{equation}
where $F^{'}=\partial F/\partial\phi$. These equations are easily
checked by substituting (\ref{4.5}) into (\ref{4.3}) and
(\ref{4.4}).

We wish to construct pure gravitational wave perturbations i.e.
having pure type N perturbed Weyl tensor in the Petrov
classification. It is shown in \cite{isotropic} that on account of
(\ref{4.6}) the $F^{'}$--parts of $E_{ab}$ and $H_{ab}$ above are
type N with degenerate principal null direction $\phi^{,a}$. Then
if we also require the $F$ parts of $E_{ab}$ and $H_{ab}$ to be
type N the perturbations we have constructed describe pure
gravitational waves with propagation direction $\phi^{,a}$ in the
Robertson--Walker background and the histories of the wave--fronts
are the null hypersurfaces $\phi(x^a)=\rm {const}$. Making use of
the following null tetrad, $k_{a}=-\dot{\phi}^{-1}\,\phi_{,a}$,
$l_{a}=u_{a}-\frac{1}{2}\,k_{a}$ and $m_{a}$, $\bar{m}_{a}$ a
complex covariant vector field and its complex conjugate chosen so
they are null ($m^a\,m_a=0=\bar{m}^a\,\bar{m}_a$), are orthogonal
to $k^a$ and $l^a$ and satisfy $m^{a}\,\bar{m}_{a}=1$ we find that
a simple way to ensure the $F$--parts of $E_{ab}$ and $H_{ab}$ are
type N is to require the null hypersurfaces
$\phi(x^{a})=\rm{const}$ to satisfy (see \cite{isotropic})
\begin{equation}\label{4.12}
\phi_{,b|c}\,\bar{m}^{b}\,l^{c}=0\ ,
\end{equation}
and
\begin{equation}\label{4.13}
\phi_{,a|b}\,m^{a}\,m^{b}=0\ .
\end{equation}

To exhibit explicit examples we specialise to the case
$\phi=x-T(t)$ with $T(t)$ introduced in (\ref{3.5}). Then the null
tetrad described above is given by the 1--forms
\begin{eqnarray}\label{4.14}
k_a\,dx^a&=&R\,dx-dt\ ,\qquad l_a\,dx^a=-\frac{1}{2}\,(R\,dx+dt)\
,
\nonumber\\
m_a\,dx^a&=&\frac{1}{\sqrt{2}}R\,p_0^{-1}f\,(dy+i\,dz)\ ,
\end{eqnarray}
and it is straightforward to check that Eqs. (\ref{4.12}) and
(\ref{4.13}) are satisfied. Since $s^{ab}$ and $\Pi^{ab}$ are
trace--free and orthogonal to $u^a$ and $\phi^{,a}$, they each
have only two independent components. These components are
$s^{22}=-s^{11}=\hat{\alpha}(y,z,x,t)$,
$s^{12}=s^{21}=\hat{\beta}(y,z,x,t)$ and
$\Pi^{22}=-\Pi^{11}=A(y,z,x,t)$, $\Pi^{12}=\Pi^{21}=B(y,z,x,t)$
where we have labelled the coordinates $x^{1}=y\ , x^{2}=z\ ,
x^{3}=x\ , x^{4}=t$. Now we can write
\begin{equation}\label{4.15}
s^{ab}=\bar s\,m^a\,m^b+s\,\bar m^a\,\bar m^b\ ,
\end{equation}
with
\begin{equation}\label{4.16}
\bar s=-R^2p_0^{-2}f^{2}(\hat{\alpha} +i\,\hat{\beta} )\ ,
\end{equation}
and
\begin{equation}\label{4.17}
\Pi ^{ab}=\bar\Pi\,m^a\,m^b+\Pi\,\bar m^a\,\bar m^b\ ,
\end{equation}
with
\begin{equation}\label{4.18}
\bar\Pi =-R^2p_0^{-2}f^2(A+i\,B)\ .
\end{equation}
It follows from (\ref{4.7}) that $\hat{\alpha}$, $\hat{\beta}$ and
$A$, $B$ must satisfy the Cauchy--Riemann equations
\begin{equation}\label{4.19}
\frac{\partial}{\partial y}(p_0^{-4}\hat{\alpha}
)-\frac{\partial}{\partial z} (p_0^{-4}\hat{\beta )}=0\ ,
\end{equation}
\begin{equation}\label{4.20}
\frac{\partial}{\partial y}(p_0^{-4}\hat{\beta}
)+\frac{\partial}{\partial z} (p_0^{-4}\hat{\alpha} )=0\ .
\end{equation}
and\begin{equation}\label{4.21} \frac{\partial}{\partial
y}(p_0^{-4}A )-\frac{\partial}{\partial z} (p_0^{-4}B )=0\ ,
\end{equation}
\begin{equation}\label{4.22}
\frac{\partial}{\partial y}(p_0^{-4}B )+\frac{\partial}{\partial
z} (p_0^{-4}A )=0\ .
\end{equation}
If we define
${\cal{G}}=p_{0}^{-4}\,f^{3}\,R^{3}(\hat{\alpha}+i\,\hat{\beta})$
and note that $f=f(x)$, $R=R(t)$ and $\hat{\alpha}$, $\hat{\beta}$
satisfy Eqs. (\ref{4.19}) and (\ref{4.20}) then ${\cal{G}}$ is an
analytic function of $\zeta:=y+iz$. We can now rewrite Eq.
(\ref{4.16}) as
\begin{equation}\label{4.23}
\bar s=-R^{-1}p_0^2f^{-1}{\cal G}(\zeta , x, t )\ .
\end{equation}
From the propagation equation (\ref{4.9}) we find
\begin{equation}\label{4.24}
\bar\Pi =-2\,R^{-2}p_0^2f^{-1}(D{\cal G}+\dot R\,{\cal G})\ ,
\end{equation}
where $D$ is given by $D=\partial /\partial x+R\,\partial /
\partial t= \partial /\partial x+\partial /\partial T$ and the dot
indicates differentiation with respect to $t$. As a consequence of
this and (\ref{4.18}) $A+iB$ is analytic in $\zeta$ and so Eqs.
(\ref{4.21}) and (\ref{4.22}) are automatically satisfied.
Replacing $s^{ab}$ by Eqs. (\ref{4.15}) and (\ref{4.23}) and
$\Pi^{ab}$ by Eqs. (\ref{4.17}) and (\ref{4.24}) the wave equation
(\ref{4.8}) simplifies to
\begin{equation}\label{4.25}
D^2{\cal G}+k\,{\cal G}=0\ ,
\end{equation}
with $k=0,\pm 1$ labelling the Robertson--Walker backgrounds with
line--elements of the form (\ref{3.4}). The solutions of these
three differential equations are: \newline for $k=0$,
\begin{equation}\label{4.26}
{\cal G}(\zeta , x, t )=a(\zeta , x-T)\,(x+T)+b(\zeta , x-T)\ ,
\end{equation}
for $k=+1$,
\begin{equation}\label{4.27}
{\cal G}(\zeta , x, t )=a(\zeta , x-T)\,\sin\left
(\frac{x+T}{2}\right )+b(\zeta , x-T)\, \cos\left
(\frac{x+T}{2}\right )\ ,
\end{equation}
and for $k=-1$,
\begin{equation}\label{4.28}
{\cal G}(\zeta , x, t )=a(\zeta , x-T)\,\sinh\left
(\frac{x+T}{2}\right )+b(\zeta , x-T)\, \cosh\left
(\frac{x+T}{2}\right )\ ,
\end{equation}
where in each case $a(\zeta , x-T) ,\ b(\zeta , x-T)$ are
arbitrary functions. Using the identity $x+T=2x-(x-T)$, (and some
simple trigonometric and hyperbolic relations) we can rewrite
(\ref{4.26}) in the form
\begin{equation} \label{4.29}
{\cal G}(\zeta, x, t)=h_1(\zeta , x-T)+xh_2(\zeta , x-T)\ ,
\end{equation}
with $h_1$, $h_2$ arbitrary, (\ref{4.27}) as
\begin{equation} \label{4.30}
{\cal G}(\zeta, x, t)=h_3(\zeta , x-T)\sin{x}+h_4(\zeta ,
x-T)\cos{x}\ ,
\end{equation}
with $h_3$, $h_4$ arbitrary and (\ref{4.28}) as
\begin{equation} \label{4.31}
{\cal G}(\zeta, x, t)=h_5(\zeta , x-T)\sinh{x}+h_6(\zeta ,
x-T)\cosh{x}\ ,
\end{equation}
with $h_5$, $h_6$ arbitrary. In addition (\ref{4.31}) can be put
in the form
\begin{equation} \label{4.32}
{\cal G}(\zeta, x, t)=h_7(\zeta , x-T)e^{x}+h_8(\zeta ,
x-T)e^{-x}\ .
\end{equation}
When these results are derived from metric perturbations in
Section 5 below the expressions (\ref{4.29})--(\ref{4.32}) will be
more useful for comparison purposes than the equivalent
expressions (\ref{4.26})--(\ref{4.28}).

The ``electric" and ``magnetic" parts of the Weyl tensor (Eqs.
(\ref{4.10}) and (\ref{4.11}) respectively) are now calculated and
we find that they can be written compactly as \cite{isotropic}
\begin{equation}\label{4.33}
E^{ab}+i\,H^{ab}=-2\,R^{-2}p_0^2f^{-1}\frac{\partial}{\partial x}
({\cal G}\,F)\,m^a\,m^b\ .
\end{equation}
Here ${\cal{G}}$ is given by Eqs. (\ref{4.26})--(\ref{4.28}) (or
equivalently (\ref{4.29})--(\ref{4.32})) and $F=F(x-T)$ so that
$F^{'}=\partial F/\partial x$, $p_0=1+(K/4)(y^2+z^2)$, $f=f(x)$
described in the previous section and $R(t)$ is the scale factor.
It follows from Eqs. (\ref{4.5}), (\ref{4.15}), (\ref{4.17}),
(\ref{4.23}) and (\ref{4.24}) that to find $\sigma_{ab}$,
$\pi_{ab}$ from $s_{ab}$ and $\Pi_{ab}$ we simply replace
${\cal{G}}$ by ${\cal{G}}\,F$. This does not affect Eqs.
(\ref{4.24}) and (\ref{4.25}) since $DF=0$. Conversely with
$F=F(x-T)$ and ${\cal{G}}$ given by Eqs.
(\ref{4.26})--(\ref{4.28}) $F$ can be absorbed into ${\cal{G}}$.

\setcounter{equation}{0}
\section{The Perturbed Metric}\indent

We now exhibit a line--element which (i) can be viewed as a
perturbation of the space--time line--element (\ref{3.4}) and (ii)
produces the same explicit perturbations described in the
gauge--invariant formalism of the previous section. We first
introduce a pair of null coordinates,
\begin{equation}\label{5.1}
u=\frac{1}{\sqrt{2}}\,(x-T(t))\ , \qquad
v=\frac{1}{\sqrt{2}}\,(x+T(t))\ ,
\end{equation}
with $T(t)$ introduced after (\ref{3.5}). Writing
\begin{equation}\label{5.2}
R(t(T)) \equiv \Omega(T)=\Omega(v-u)\ ,
\end{equation}
the line--element (\ref{3.4}) written in terms of $u$ and $v$
reads
\begin{equation}\label{5.3}
ds^{2}=\Omega^{2}\,{p_{0}^{-2}\,f^2(dy^2+dz^2)} +
2\,\Omega^2\,du\,dv\ ,
\end{equation}
where now $f=f(u+v)$. The coordinates ${y,z,u,v}$ are such that
the surfaces $u=\rm{const}$, $v=\rm{const}$ are two families of
intersecting null hypersurfaces. The general form of line--element
in a coordinate system based upon two families of intersecting
null hypersurfaces is given in \cite{israel}. For our purposes we
write this as
\begin{equation}\label{5.4}
ds^2=b^2\,h_{AB}\,(dx^A+a_{1}^{A}\,du+a_{2}^{A}\,dv)(dx^B+a_{1}^{B}\,du+a_{2}^{B}\,dv)
+ 2\,c\,du\,dv\ ,
\end{equation}
where $A$, $B$ take values $(1,2)$, $(h_{AB}(y,z,u,v))$ is a
unimodular $2\times2$ symmetric matrix, $(x^{1}, x^{2})=(y,z)$ and
$a_{1}^{A}$, $a_{2}^{A}$, $b$, $c$ are six functions of
${y,z,u,v}$. It is convenient to use the following parametrisation
\cite{H+T} of $(h_{AB})$:
\begin{equation}\label{5.5}
(h_{AB})=\left( \begin{array}{cc} e^{2\alpha}\,\cosh2\beta & \sinh2\beta\\
\sinh2\beta & e^{-2\alpha}\,\cosh2\beta \end{array} \right)\ .
\end{equation}
Here $\alpha$, $\beta$ are taken to be small of first order. With
$(h_{AB})$ given by Eq. (\ref{5.5}) it is easy to check that,
working to first order, Eq. (\ref{5.4}) can be written
\begin{eqnarray}\label{5.6}
ds^2&=&b^2[(1+\alpha)dy+\beta\,
dz+\{a^1_1(1+\alpha)+a^2_{1}\,\beta\}\,du+\{a^1_2(1+\alpha)+a^2_2\,\beta\}dv]^2
\nonumber \\& &
 +b^2[\beta\, dy+(1-\alpha)
dz+\{a^1_1\,\beta+a^2_{1}(1-\alpha)\}\,du+\{a^1_2\,\beta+a^2_2(1-\alpha)\}dv]^2
 \nonumber \\ & & +2\,c\,du\,dv \ .
\end{eqnarray}
The background space--time is obtained from this by putting
\begin{equation}\label{5.7}
a_1^A=0\ , \qquad a_2^A=0\ , \qquad b=p_0^{-1}\,\Omega\,f\ ,
\qquad c=\Omega ^2\ , \qquad  \alpha = 0\ , \qquad \beta=0\ .
\end{equation}
For the perturbed space--time that we require we find that $b$,
$c$ retain their background values, and we can put
$a_1^A=0=a_2^A$. These latter quantities actually play the role of
gauge terms (see Section 7 below and Appendix B for an
illustration of this). Every shear--free system of gravitational
waves involves an arbitrary analytic function \cite{Rob} and we
now have two real functions $\alpha$, $\beta$ available to provide
the real and imaginary parts of this analytic function. Also we
find that the 4--velocity $u^a$, the isotropic pressure $p$ and
the matter--energy density $\mu$ take their background values
(these are given in Section 3).

To demonstrate that this space--time does indeed describe the
perturbations of Section 4 we shall work on the tetrad given via
the 1--forms
\begin{eqnarray}\label{5.8}
\theta^1 &=& p_0^{-1}\,f\,\Omega\,\{(1+\alpha)\,dy+\beta\,dz\}\ ,
\nonumber \\
\theta^2 &=& p_0^{-1}\,f\,\Omega\,\{\beta\,dy+(1-\alpha)\,dz\}\ ,
\nonumber \\
\theta^3 &=& \Omega\,du\ , \nonumber \\
\theta^4 &=& \Omega\,dv\ ,
\end{eqnarray}
with $p_0=1+(K/4)(y^2+z^2)$ as in (\ref{3.4}). We note that with
respect to this tetrad the line--element is now
\begin{equation}\label{5.9}
ds^2=(\theta^1)^2+(\theta^2)^2+2\,\theta^3\,\theta^4=g_{ab}\,\theta^a\,\theta^b\
,
\end{equation}
thus defining the tetrad components $g_{ab}$ of the metric tensor.
The tetrad components of the matter 4--velocity are given via the
1--form
\begin{equation}\label{5.10}
u_a\,\theta^a=\frac{1}{\sqrt{2}}\,(\theta^3-\theta^4)\ .
\end{equation}
Since we wish to reproduce the linear perturbations of the
previous section we shall discard any terms which are second order
or smaller in $\alpha$ and $\beta$. Our first step is to calculate
the Ricci rotation coefficients and the Ricci tensor components.
This results in a lengthy list of equations which for convenience
we give in Appendix A. We now use the Ricci tensor components and
(\ref{5.10}) in the field equations given by Eqs. (\ref{2.11}).
Noting that
\begin{equation}\label{5.11}
\Omega^{'}=\frac{1}{\sqrt{2}}\,R\,\dot{R}\ , \qquad
\Omega^{''}=\frac{1}{2}\,R^{2}\,\ddot{R}+\frac{1}{2}\,R\,\dot{R}^{2}\
,
\end{equation}
where the prime denotes differentiation with respect to $v$ and
using Eqs. (\ref{3.2}) and (\ref{3.3}) it is easily checked that
the first of Eqs. (\ref{2.11}) is identically satisfied. The
second equation in (\ref{2.11}) yields
\begin{eqnarray}\label{5.12}
q_{1} &=&
\frac{p_{0}^3}{\sqrt{2}}\,f^{-1}\Omega^{-2}\{(p_{0}^{-2}\alpha)_{yu}+(p_{0}^{-2}\beta)_{zu}
-(p_{0}^{-2}\alpha)_{yv}-(p_{0}^{-2}\beta)_{zv}\} \ , \\ q_{2}&=&
\frac{p_{0}^3}{\sqrt{2}}\,f^{-1}\Omega^{-2}\{(p_{0}^{-2}\alpha)_{zv}-(p_{0}^{-2}\beta)_{yv}
-(p_{0}^{-2}\alpha)_{zu}+(p_{0}^{-2}\beta)_{yu}\} \ ,
\\
q_{3}&=&0\ , \\
q_{4}&=&0\ .
\end{eqnarray}
The subscripts $y,z,u,v$ here indicate partial differentiation
with respect to these variables. We recall that in the covariant
approach we found that $q_{a}\equiv 0$. With $\alpha$, $\beta$
chosen so they satisfy the Cauchy--Riemann equations in the form:
\begin{equation}\label{5.16}
(p_{0}^{-2}\,\alpha)_{y}+(p_{0}^{-2}\,\beta)_{z} = 0 \ ,
\end{equation}
\begin{equation}\label{5.17}
(p_{0}^{-2}\,\alpha)_{z}-(p_{0}^{-2}\,\beta)_{y} = 0 \ ,
\end{equation}
it follows that $q_{a}\equiv 0$ as required. For later use we note
that as a result of these equations $p_0^{-2}(\alpha-i\,\beta)$ is
analytic in $\zeta=y+iz$. With $q_a=0$ the last of Eqs.
(\ref{2.11}) can be rewritten as
\begin{equation}\label{5.18}
R_{ab}=\mu\,u_{a}u_{b}+p\,h_{ab}+\pi_{ab}-\frac{1}{2}(3p-\mu)\,g_{ab}\
.
\end{equation}
With the Ricci tensor components given by (A.11)--(A.20) it
follows from this and Eqs. (\ref{3.2}), (\ref{3.3}), (\ref{5.9})--
(\ref{5.11}), (\ref{5.16}), (\ref{5.17}) that $\pi_{ab}=0$ except
for $\pi_{11}$, $\pi_{22}$ and $\pi_{12}$ with
\begin{eqnarray}\label{5.19}
\pi_{11}&=&\sqrt{2}\,R^{-2}\,\dot{R}(\alpha_v-\alpha_u)-2\,R^{-2}\,f^{-1}\,f^{'}(\alpha_v+\alpha_u)-2\,R^{-2}\,\alpha_{vu}\
, \\
\pi_{22}&=&-\sqrt{2}\,R^{-2}\,\dot{R}(\alpha_v-\alpha_u)+2\,R^{-2}\,f^{-1}\,f^{'}(\alpha_v+\alpha_u)+2\,R^{-2}\,\alpha_{vu}\
, \\
\pi_{12}&=&\sqrt{2}\,R^{-2}\,\dot{R}(\beta_v-\beta_u)-2\,R^{-2}\,f^{-1}\,f^{'}(\beta_v+\beta_u)-2\,R^{-2}\,\beta_{vu}\
.
\end{eqnarray}
We have made use of
\begin{equation}\label{5.22}
2f^{''}=-k\,f\ , \qquad 2(f^{'})^{2}+k\,f^2=K\ ,
\end{equation}
to simplify these equations. We note that the prime here denotes
differentiation with respect to $v=(1/\sqrt{2})(x+T)$. Similar
equations to these appear in \cite{isotropic} (eq. (5.41)). In
\cite{isotropic} the prime indicates differentiation with respect
to $x$ and hence the factors of $2$ in (\ref{5.22}) do not appear
there.

 Now in terms of the background null tetrad described by Eq. (\ref{4.14})
 we can write the coordinate components of the (small) anisotropic
 stress tensor as
 \begin{equation}\label{5.23}
 \pi^{ij}=\bar{\pi}\,m^i\,m^j+\pi\,\bar{m}^i\,\bar{m}^j\ .
\end{equation}
Using Eqs. (\ref{4.14}) and (\ref{5.8}) (with $\alpha=\beta=0$
since $\pi_{ab}$ is a first order quantity) we find
\begin{equation}\label{5.24}
\bar{\pi}=\frac{1}{2}(\pi_{11}-\pi_{22})-i\pi_{12}\ .
\end{equation}
Substituting from Eqs. (\ref{5.19})--(5.21) above yields
\begin{eqnarray}\label{5.25}
\bar{\pi}&=&-2\,p_{0}^{2}\,f^{-1}\,R^{-2}\,[p_{0}^{-2}\,f(\alpha-i\beta)_{uv}+p_{0}^{-2}\,f^{'}\{\alpha_{v}+\alpha_{u}-i(\beta_{v}
+\beta_{u})\}] \nonumber \\ & & -
\sqrt{2}R^{-2}\,\dot{R}\{\alpha_{u}-\alpha_{v}-i(\beta_{u}-\beta_{v})\}\
.
\end{eqnarray}
In the previous section we worked with $\Pi_{ab}$ and $\bar{\Pi}$
which we expressed in terms of an analytic function $\cal{G}$. But
$\pi_{ab}=\Pi_{ab}\,F$ and as indicated following Eq. (\ref{4.33})
$F$ can be absorbed into $\cal{G}$. Hence, in order to make
contact with the gauge--invariant description $\bar{\pi}$ here
must satisfy the same equation as $\bar{\Pi}$ and thus we require
\begin{equation}\label{5.26}
\bar{\pi}=-2\,p_{0}^{2}\,f^{-1}\,R^{-2}\{D{\cal{G}}+\dot{R}\,{\cal{G}}\}\
,
\end{equation}
with
$D=\partial/\partial{x}+R\partial/\partial{t}=\sqrt{2}\,\partial/\partial{v}$
for some analytic function $\cal{G}$. Taking
\begin{equation}\label{5.27}
{\cal{G}}=\frac{1}{\sqrt{2}}\,p_{0}^{-2}\,f\{\alpha_{u}-\alpha_{v}-i(\beta_{u}-\beta_{v})\}\
,
\end{equation}
we find that it is indeed possible to write $\bar{\pi}$ in this
form provided we choose $\alpha$, $\beta$ to satisfy the
following:
\begin{equation}\label{5.28}
\mbox{If }f^{'}=0 \mbox{ then } \alpha_{vv}=0\ , \beta_{vv}=0\ ;
\end{equation}
\begin{equation}\label{5.29}
\mbox{if }f^{'}\neq 0 \mbox{ then } \alpha_{v}=0\ , \beta_{v}=0\ .
\end{equation}
We note that the first of these conditions corresponds to the case
$k=0$, $K=0$ described following Eq. (\ref{3.4}). We now assume
that these conditions hold. As a consequence of these and Eq.
(\ref{5.22}) it immediately follows that $\cal{G}$ given by
(\ref{5.27}) satisfies the wave equation (\ref{4.25}). Also noting
that $f=f(x)$ and using the Cauchy--Riemann equations
(\ref{5.16})--(\ref{5.17}) we see that as before $\cal{G}$ is an
analytic function of $\zeta=y+iz$.

We now turn our attention to the shear. In a similar fashion to
the anisotropic stress the coordinate components of the (small)
shear tensor can be written in the form
\begin{equation}\label{5.30}
\sigma_{ij}=\bar{s}_p\,m_{i}\,m_{j}+s_{p}\,\bar{m}_i\,\bar{m}_j\ ,
\end{equation}
where $m_i$, $\bar{m}_j$ are given by Eq. (\ref{4.14}) and in
terms of the Ricci rotation coefficients
\begin{equation}\label{5.31}
\bar{s}_p=\frac{1}{2\sqrt{2}}\{(\Upsilon_{141}-\Upsilon_{242}-\Upsilon_{131}+\Upsilon_{232})+i(\Upsilon_{132}+\Upsilon_{231}
-\Upsilon_{142}-\Upsilon_{241})\}\ .
\end{equation}
Evaluating this using the Ricci rotation coefficients given in
Appendix A we find
\begin{equation}\label{5.32}
\bar{s}_p=\frac{1}{\sqrt{2}}\,R^{-1}\{(\alpha-i\beta)_{v}-(\alpha-i\beta)_u\}=-p_{0}^{-2}\,f^{-1}\,R^{-1}\,{\cal{G}}\
,
\end{equation}
with ${\cal{G}}$ as before. Taking into account that we can absorb
$F$ into ${\cal{G}}$ and that $\bar{s}_p=\bar{s}\,F$ (with
$\bar{s}$ defined by Eq. (\ref{4.15})) we see that the
perturbations we have produced here also satisfy Eq. (\ref{4.23}).
Thus we have shown that \emph{the perturbations described by the
metric} (\ref{5.6}) \emph{take the same form as those found by the
covariant approach}. In the next section we shall illustrate that
they also satisfy the wave equation (\ref{4.8}) and the
propagation equation (\ref{4.9}).

For the remainder of this section we compare the explicit
${\cal{G}}$ found here with the solutions of the wave equation
found in \cite{isotropic} and listed in Eqs.
(\ref{4.26})--(\ref{4.28}) (or equivalently
(\ref{4.29})--(\ref{4.32})). \emph{We first examine the case when
$k=0$}. There are two subcases to consider here (i) $K=0$ and
$f(x)=1$, (ii) $K=+1$ and $f(x)=x$. When $K=0$, $p_{0}=1$ and Eq.
(\ref{5.27}) reads
\begin{equation}\label{5.33}
{\cal{G}}=\frac{1}{\sqrt{2}}\{(\alpha-i\beta)_u-(\alpha-i\beta)_v\}\
.
\end{equation}
Since $f^{'}=0$ in this case we have $\alpha_{vv}=0=\beta_{vv}$.
Thus in addition to $\alpha-i\beta$ being analytic in $\zeta$ this
complex--valued function is also linear in $v$. Hence we can write
\begin{equation}\label{5.34}
{\cal{G}}(\zeta,\, x,\, t) = a_{1}(\zeta, \, x-T)(x+T)
+a_{2}(\zeta, \, x-T)\ ,
\end{equation}
where $a_1$, $a_2$ are arbitrary (analytic) functions of their
arguments. When $K=+1$, $f(x) = 0$ and from (\ref{5.29}) we have
$\alpha_v=0=\beta_v$. Therefore the function
$p_{0}^{-2}(\alpha-i\beta)$ is analytic in $\zeta$ and independent
of $v=(x+T)/\sqrt{2}$, i.e. it depends only on $\zeta$ and
$u=(x-T)/\sqrt{2}$, and we can write (\ref{5.27}) in the form
\begin{equation}\label{5.35}
{\cal{G}}=x\,a_3(\zeta,\, x-T)\ .
\end{equation}
with $a_3$ an arbitrary analytic function. Using the identity
$x+T=2x-(x-T)$ as in Section 4 we can rewrite (\ref{5.34}) in the
form (\ref{4.29}). Then (\ref{5.35}) is the special case of
(\ref{4.29}) corresponding to $h_1(\zeta, x-T)\equiv 0$.
\emph{Thus in the case $k=0$ there are two independent expressions
for ${\cal G}(\zeta, x, t)$ which are given in the form of a
superposition in} (\ref{4.29}). This arises because (\ref{4.29})
is obtained by solving the linear wave equation (\ref{4.25}) with
$k=0$ and in general this equation is insensitive to the allowable
values of $K=0, \pm1$.

We now look at the solution when $k=+1$. There is only one case to
consider here, $K=+1$ with $f(x)=\sin{x}$ or equivalently (see
Section 3) $f(x)=\cos{x}$. Again we have
$p_{0}^{-2}(\alpha-i\beta)$ analytic in $\zeta$ and independent of
$v$ so (\ref{5.27}) can now be written
\begin{equation}\label{5.36}
{\cal{G}}=a_4(\zeta, \, x-T)\,\sin{x}\  \mbox{  or }\
{\cal{G}}=a_5(\zeta, \, x-T)\,\cos{x}\ ,
\end{equation}
where $a_4$, $a_5$ are arbitrary functions. \emph{The two
equations in} (\ref{5.36}) \emph{are equivalent to} (\ref{4.30}).

Finally when $k=-1$ there are three subcases to look at
corresponding to $K=0,\pm1$. In all cases $f^{'}(x) \neq 0$ and so
we have $p_{0}^{-2}(\alpha-i\beta)$ independent of $v$ and
${\cal{G}}$ has the form
\begin{equation}\label{5.37}
{\cal{G}}=f(x)\,a_{6}(\zeta,\, x-T)\ ,
\end{equation}
where $a_6$ is an arbitrary analytic function. When $K=0$ we have
$f(x)=\frac{1}{2}e^x$ or equivalently $f(x)=\frac{1}{2}e^{-x}$ and
so \emph{in this case} (\ref{5.37}) \emph{agrees with}
(\ref{4.32}). When $K=+1$, $f(x)=\sinh{x}$ and now (\ref{5.37})
\emph{agrees with} (\ref{4.31}) when $h_6(\zeta, x-T)\equiv0$.
When $K=-1$, $f(x)=\cosh{x}$ and (\ref{5.37}) \emph{agrees with}
(\ref{4.31}) when $h_5(\zeta, x-T)\equiv0$. This case $k=-1$ is a
good illustration of the insensitivity of the expressions
(\ref{4.31}) and (\ref{4.32}) to the values of $K$.

Thus all of the solutions found here are identical to the
solutions found using the gauge--invariant and covariant approach
to perturbations in \cite{isotropic}.

\setcounter{equation}{0}
\section{Properties of the Shear and Anisotropic Stress}\indent
In the previous section we exhibited a perturbation of the
Robertson--Walker background line--element (\ref{3.4}) that
produced perturbations in the shear and anisotropic stress tensors
which satisfied some of the equations found using the
gauge--invariant and covariant approach of Section 4. We now show
that these perturbations satisfy the remaining equations, namely
that the anisotropic stress and shear tensors are trace--free,
orthogonal to $u^a$, divergence--free with respect to the
background metric and also satisfy the wave equation (\ref{4.8})
and propagation equation (\ref{4.9}). To do this we shall, in this
section, work in coordinate components [in the local coordinates
${y,z,x,t}$] instead of the tetrad components we have used up to
this point. In terms of this local coordinate system we can write
the line--element (\ref{5.6}) (with $a_1^A=a_2^A=0$,
$b=p_0^{-1}\Omega\,f$, $c=\Omega ^2$) in the form
\begin{equation}\label{6.1}
ds^2=\hat{g}_{ab}\,dx^a\,dx^b+2\gamma_{ab}\,dx^a\,dx^b:=g_{ab}\,dx^a\,dx^b\
,
\end{equation}
where $\hat{g}_{ab}= {\rm{diag}}\{p_{0}^{-2}\,f^2\,\Omega^{2},
p_{0}^{-2}\,f^2\,\Omega^{2}, \Omega^{2}, -1\}$ is the metric of
the background space--time and
\begin{equation}\label{6.2}
\gamma_{ab}=p_0^{-2}\,f^2\,\Omega^{2}\left(\begin{array}{cccc} \alpha&\beta&0&0\\
\beta&-\alpha&0&0\\ 0&0&0&0\\ 0&0&0&0
\end{array}\right)
\end{equation}
is the perturbation. Clearly $\gamma_{ab}$ is trace--free and
orthogonal to $u^{a}=\delta^a_4$ (with
$\hat{g}_{ab}\,u^a\,u^b=-1$). The non--vanishing Christoffel
symbols of the background metric tensor given via the
line--element (\ref{3.4}) are:
\begin{eqnarray}\label{6.3}
\hat{\Gamma}^1_{11} &=& -\hat{\Gamma}^1_{22}=\hat{\Gamma}^2_{12}=-\frac{1}{2}\,K\,p_0^{-1}\,y \ , \nonumber \\
\hat{\Gamma}^1_{12}&=&-\hat{\Gamma}^2_{11}=\hat{\Gamma}^2_{22}=-\frac{1}{2}\,K\,p_0^{-1}\,z \ , \nonumber \\
\hat{\Gamma}^1_{14} &=&\hat{\Gamma}^2_{24}=\hat{\Gamma}^3_{34}=\Omega^{-1}\,\Omega_{t}\ , \nonumber \\
\hat{\Gamma}^4_{11} &=&
\hat{\Gamma}^4_{22}=p_{0}^{-2}\,f^{2}\,\Omega\,\Omega_{t}\ ,
\\
\hat{\Gamma}^3_{11}&=& \hat{\Gamma}^3_{22}=-p_{0}^{-2}\,f\,f_{x}\
,
\nonumber \\
\hat{\Gamma}^1_{13}&=& \hat{\Gamma}^2_{23}=f^{-1}\,f_{x}\ ,
\nonumber \\
\hat{\Gamma}^4_{33}&=& \Omega\,\Omega_{t}\ . \nonumber
\end{eqnarray}
We have used the hat here to emphasise that these are background
Christoffel symbols and we shall continue to use this notation to
denote background quantities for the remainder of this section.
Using these and the Cauchy--Riemann equations
(\ref{5.16})--(\ref{5.17}) it is a simple exercise to show that
$\gamma_{ab}$ defined above is divergence--free.

In order to show that ${\pi}^{ab}$ is also divergence--free we
first write it in terms of $\gamma_{ab}$. We define the
perturbation of the Christoffel symbols to be
$\delta\,\Gamma^a_{bd}:=\Gamma^a_{bd}-\hat{\Gamma}^a_{bd}$. Noting
that $g^{ab}=\hat{g}^{ab}-\gamma^{ab}$ (here
$\gamma^{ab}=\hat{g}^{ac}\,\hat{g}^{bf}\,\gamma_{cf}$ and we are
neglecting second order small quantities ) it is easily derived
from the definition of the Christoffel symbols that
\begin{equation}\label{6.4}
\delta\,\Gamma^a_{bd}=\frac{1}{2}\,(\gamma^a_{b|d}+\gamma^a_{d|b}-\hat{g}^{af}\,\gamma_{bd|f})\
,
\end{equation}
where as usual the stroke indicates differentiation with respect
to the background metric. Now $\gamma^{ab}$ is divergence--free
and thus we can see from this equation that
$\delta\,\Gamma^a_{ba}=0$. In general the components of the Ricci
tensor of a perturbed metric can be written in the form
\begin{equation}\label{6.5}
R_{bd}=\hat{R}_{bd}+(\delta\,\Gamma^a_{bd})_{|a}-(\delta\,\Gamma^a_{ba})_{|d}\
.
\end{equation}
For the problem at hand we have
\begin{equation}\label{6.6}
\hat{R}_{bd}=\mu\,u_b\,u_d+p\,\hat{h}_{bd}-\frac{1}{2}\,(3p-\mu)\,\hat{g}_{bd}\
.
\end{equation}
Substituting for $R_{bd}$ and $\hat{R}_{bd}$ from Eqs.
(\ref{5.18}) and (\ref{6.6}) respectively in Eq. (\ref{6.5})
yields
\begin{equation}\label{6.7}
(\delta\,\Gamma^a_{bd})_{|a}=\pi_{bd}+\frac{1}{2}\,(\mu-p)\gamma_{bd}\
.
\end{equation}
Taking the divergence of Eq. (\ref{6.4}) and using this equation
we arrive at
\begin{equation}\label{6.8}
\gamma^a_{b|da}+\gamma^a_{d|ba}-\hat{g}^{af}\gamma_{bd|fa}=2\pi_{bd}+(\mu-p)\,\gamma_{bd}\
.
\end{equation}
Next making use of the Ricci identities
\begin{equation}\label{6.9}
\gamma_{ab|dc}-\gamma_{ab|cd}=R_{afcd}\,\gamma^f_b-R_{fbcd}\,\gamma^f_a\
,
\end{equation}
and recalling that $\gamma^{ab}$ is divergenceless and orthogonal
to $u^a$ we find (since $C_{abcd}=0$ in the background)
\begin{equation}\label{6.10}
\gamma^c_{a|dc}=\frac{3}{2}\,\gamma^c_a\,R_{cd}+\frac{1}{2}\,\gamma^c_d\,R_{ac}-\frac{1}{2}\,g_{ad}\,R_{fc}\,\gamma^{fc}
-\frac{1}{6}\,R\,\gamma_{ad}\ .
\end{equation}
With $R=\mu-3p$ and $R_{ab}$ given by Eq. (\ref{5.18}) this
equation allows us to write (since we are concerned here with
first--order terms only)
\begin{equation}\label{6.11}
\gamma^a_{b|da}+\gamma^a_{d|ba}=\left(\frac{5}{3}\,\mu-p\right)\,\gamma_{bd}\
,
\end{equation}
and hence Eq. (\ref{6.8}) now becomes
\begin{equation}\label{6.12}
\hat{g}^{af}\gamma_{bd|fa}-\frac{2}{3}\,\mu\,\gamma_{bd}=-2\,\pi_{bd}\
.
\end{equation}
It is easy to see from this that $\pi_{ab}$ is trace--free and
orthogonal to $u^a$. Starting with this equation we shall now
prove that $\pi_{ab}$ is indeed divergence--free. That this is
necessary to fully make contact with the gauge--invariant and
covariant approach of Section 4 follows from the fact that in this
case we wrote $\pi_{ab}=\Pi_{ab}\,F$ with $\Pi^{ab}{}_{|b}=0$,
$F=F(x-T)$ and $\Pi^{ab}=0$ except for $\Pi^{11}$, $\Pi^{22}$ and
$\Pi^{12}$ and therefore $\Pi^{ab}{}_{|b}=0$ is equivalent to
$\pi^{ab}{}_{|b}=0$ in this case. First making use of the Ricci
identities for a tensor of type (3,0), Eq. (\ref{2.1}) and Eq.
(\ref{2.10}) we can write
\begin{equation}\label{6.13}
\gamma^{ab|d}{}_{|db}=(\gamma^{ab}{}_{|b})^{|d}{}_{|d}+\left(\frac{7}{6}\,\mu-\frac{1}{2}\,p\right)\,\gamma^{ad}{}_{|d}=0\
.
\end{equation}
Also since for the perturbed space--time we are considering here
the matter density $\mu$ retains its background value we have
$h^b_c\,\mu_{,b}=0$ from which it follows that
\begin{equation}\label{6.14}
\mu_{,b}=-\dot{\mu}\,u_b\ .
\end{equation}
As a consequence of these last two equations we find, on taking
the divergence of Eq. (\ref{6.12}), that
\begin{equation}\label{6.15}
\pi^{ab}{}_{|b}=0\ ,
\end{equation}
as required.

We shall now examine the properties of the shear $\sigma_{ab}$. As
with the anisotropic stress above it is necessary to express this
in terms of $\gamma^{ab}$. This is easily done using the
definition of the covariant derivative of $u_{a}$:
\begin{equation}\label{6.16}
u_{a;b}:=-\Gamma^{c}_{ab}\,u_c=-\hat{\Gamma}^c_{ab}\,u_c-\delta\,\Gamma^c_{ab}u_c\
.
\end{equation}
We remind the reader that the semicolon here indicates covariant
differentiation with respect to the perturbed metric (background
plus a small perturbation) while a stroke denotes covariant
differentiation with respect to the background metric. In the
background Robertson--Walker space--time the shear, vorticity and
the $4$--acceleration all vanish and so Eq. (\ref{2.5})
specialises to
\begin{equation}\label{6.17}
\hat{\Gamma}^c_{ab}\,u_c:=u_{a|b}=\frac{1}{3}\,\theta\,\hat{h}_{ab}\
,
\end{equation}
in this case. Making use of this equation and Eq. (\ref{6.4}) in
Eq. (\ref{6.16}) it follows, on account of
$h_{ab}=\hat{h}_{ab}+\gamma_{ab}$ (since for the problem at hand
$u^a$ is unperturbed), that
\begin{equation}\label{6.18}
u_{a;b}=\frac{1}{3}\,\theta\,h_{ab}+\frac{1}{2}\,\dot{\gamma}_{ab}\
.
\end{equation}
Here and for the remainder of this section a dot indicates
covariant differentiation with respect to the background metric in
the direction of $u^a$. Recalling that the $4$--acceleration is
zero in the background Robertson--Walker space--time (i.e.
$u_{a|b}u^b=0$) it is trivial to see from the latter equation that
the $4$--acceleration in the perturbed space--time also vanishes.
We also note that this equation is symmetric in $(a,b)$ and thus
it is clear from Eq. (\ref{2.6}) that, as in the covariant
approach, the vorticity tensor vanishes in the perturbed
space--time. Now equating Eqs. (\ref{6.18}) and(\ref{2.5}) with
the $4$--acceleration and vorticity tensor both zero we arrive at
a simple relationship between $\sigma_{ab}$ and $\gamma_{ab}$
namely,
\begin{equation}\label{6.19}
\sigma_{ab}=\frac{1}{2}\dot{\gamma}_{ab}\ .
\end{equation}
Using this and the properties of $\gamma_{ab}$ it is
straightforward to check that $\sigma_{ab}$ is trace--free and
orthogonal to $u^a$. However further calculation is necessary to
show that it is also divergence--free (this is required for
similar reasons to those given above while discussing the
anisotropic stress). First using the Ricci identities given in Eq.
(\ref{6.9}) and noting that $C_{abcd}=0$ we calculate
\begin{equation}\label{6.20}
\dot{\gamma}^{ab}{}_{|b}=(\gamma^{ab}{}_{|b})^{.}+\frac{3}{2}\,\gamma^{af}\,\hat{R}_{fc}\,u^c-\frac{1}{2}\,\hat{R}_{bf}\,\gamma^{bf}\,u^a
\ .
\end{equation}
Replacing $\hat{R}_{ab}$ here by the right--hand side of Eq.
(\ref{6.6}) and keeping in mind that $\gamma^{ab}{}_{|b}=0$,
$\gamma^{ab}\,u_{b}=0$ leads to
\begin{equation}\label{6.21}
\dot{\gamma}^{ab}{}_{|b}=0\ ,
\end{equation}
and therefore as a result of Eq. (\ref{6.19})
$\sigma^{ab}{}_{|b}=0$.

At this point all that remains to fully make contact with the
gauge--invariant and covariant description of gravitational wave
perturbations outlined in Section 4 is to reconstruct the wave
equation (\ref{4.8}) and the propagation equation (\ref{4.9}).
This is done as follows: Using the Ricci identities the covariant
derivative in the direction of $u^a$ of Eq. (\ref{6.12}) can be
written as
\begin{eqnarray}\label{6.22}
-2\,\dot{\pi}^{ab}&=&\gamma^{ab|d}{}_{|cd}\,u^c-\frac{1}{3}\,\theta\,\left(p+\frac{1}{3}\,\mu\right)\,\gamma^{ab}+
\frac{1}{2}(\mu+3p)\,\dot{\gamma}^{ab} \nonumber \\
&&-
\frac{2}{3}\dot{\mu}\,\gamma^{ab}-\frac{2}{3}\,\mu\,\dot{\gamma}^{ab}\
.
\end{eqnarray}
Also with $h^b_c\,\mu_{,b}=0$, $h^b_c\,p_{,b}=0$, and
$h^b_c\,\theta_{,b}=0$ we find, again using the Ricci identities
and Eq. (\ref{6.12}), that
\begin{eqnarray}\label{6.23}
\gamma^{ab|d}{}_{|cd}\,u^c&=&(\dot{\gamma}^{ab})^{|d}{}_{|d}-\frac{2}{3}\,\theta\,\left(\frac{1}{6}\,\mu+\frac{1}{2}\,p\right)\gamma^{ab}
-\frac{1}{3}\,\theta^{2}\,\dot{\gamma}^{ab} \nonumber \\
&
&-\frac{2}{3}\,\theta\left(\frac{2}{3}\,\mu\,\gamma^{ab}-2\,\pi^{ab}\right)-\frac{2}{3}\,\theta\,\ddot{\gamma}^{ab}\
.
\end{eqnarray}
Entering this into Eq. (\ref{6.22}) and replacing
$\dot{\gamma}^{ab}$ by $2\,\sigma^{ab}$ we arrive at
\begin{equation}\label{6.24}
-\dot{\pi}^{ab}-\frac{2}{3}\,\theta\,\pi^{ab}=\sigma^{ab|d}{}_{|d}-\frac{2}{3}\,\theta\,\dot{\sigma}^{ab}-\left(\frac{1}{3}
\,\dot{\theta}+\frac{4}{9}\,\theta^{2}\right)\,\sigma^{ab}+\left(p-\frac{1}{3}\,\mu\right)\,\sigma^{ab}\
.
\end{equation}
We have made use of the background values of $\dot{\theta}$ and
$\dot{\mu}$ to write the equation in this form. The background
value of $\dot{\theta}$ is
\begin{equation}\label{6.25}
\dot{\theta}=-\frac{1}{3}\,\theta^{2}-\frac{1}{2}(\mu+3\,p)
\end{equation}
which is obtained by specialising Raychaudhuri's equation to the
background (i.e. putting $q_{a}$, $\pi_{ab}$, $\dot{u}_a$,
$\sigma_{ab}$ and $\omega_{ab}$ all equal to zero) and the
background value of $\dot{\mu}$ is given by
\begin{equation}\label{6.26}
\dot{\mu}=-\theta\,(\mu+p)\ .
\end{equation}
This is found by specialising to the background the projections
along and orthogonal to $u^a$ of the conservation equation
$T^{ab}{}_{|b}=0$ (see for example Eqs. (2.20) and (2.21) in
\cite{isotropic}). Both the wave equation and the propagation
equation are actually contained in Eq. (\ref{6.24}). To confirm
this we again put
\begin{equation}\label{6.27}
\sigma_{ab}=s_{ab}\,F(\phi)\ , \qquad \pi_{ab}=\Pi_{ab}\,F(\phi)\
,
\end{equation}
where $F(\phi)$ is an arbitrary analytic function of its argument
$\phi=x-T(t)$. In the covariant approach we found
$s^{ab}\,\phi_{,b}=0$ and $\Pi^{ab}\,\phi_{,b}=0$. This is also
true here since $\phi_{,b}=(0,\,0,\,1,\,-R^{-1})$ and we have
$\pi^{3b}=0$, $\pi^{4b}=0$, $\sigma^{3b}=0$ and $\sigma^{4b}=0$.
In addition since the hypersurfaces $\phi(x^a)=\rm{const}$ are
null we have $\phi^{,d}\phi_{,d}=0$. Thus we can write
\begin{equation}\label{6.28}
\dot{\pi}^{ab}=\dot{\Pi}^{ab}\,F+\dot{\phi}\,\Pi^{ab}\,F^{'}\ ,
\end{equation}
\begin{equation}\label{6.29}
\dot{\sigma}^{ab}=\dot{s}^{ab}\,F+\dot{\phi}\,s^{ab}\,F^{'}\ ,
\end{equation}
and
\begin{equation}\label{6.30}
\sigma^{ab|d}{}_{|d}=s^{ab|d}{}_{|d}\,F+(2\,s^{ab|d}\,\phi_{,d}+s^{ab}\,\phi^{,d}{}_{|d})\,F^{'}\
,
\end{equation}
where $F^{'}=dF/d\phi$. Substituting these expressions for
$\sigma^{ab}$ and $\pi^{ab}$ into Eq. (\ref{6.24}) and equating
the $F$ and $F^{'}$ parts separately yields the required wave
equation (\ref{4.8}) and propagation equation (\ref{4.9}).

\setcounter{equation}{0}
\section{Discussion}\indent
We have shown in Sections 5 and 6 that the perturbations of the
background Robertson--Walker space--time derived here from metric
perturbations are exactly the same as those obtained using the
covariant approach. Thus the metric (\ref{5.6}) with
$a^{A}_{1}=0$, $a^{A}_{2}=0$ and $\alpha$, $\beta$ chosen to
satisfy the Cauchy--Riemann equations (\ref{5.16})--(\ref{5.17})
is indeed that which we set out to find. We mentioned earlier that
the functions $a^{A}_{1}$, $a^{A}_{2}$ play the role of gauge
terms. That this is true is seen by repeating the calculation of
${\cal{G}}$ with $a^{A}_{1}\neq0$, $a^{A}_{2}\neq0$. To save
repetition here this calculation is outlined briefly in Appendix
B. The result is that $a^{A}_{1}$, $a^{A}_{2}$ do not appear in
the required analytic function ${\cal{G}}$ i.e that which
satisfies Eq. (\ref{5.26}). Thus since all gauge invariant
perturbations can be written in terms of this ${\cal{G}}$ we
conclude that $a^{A}_{1}$, $a^{A}_{2}$ are pure gauge terms which
we can put equal to zero without loss of generality.

Metric perturbations of Robertson--Walker space--times, which can
be viewed as describing gravitational radiation, have also been
studied by Bardeen \cite{bardeen} in an important paper. In this
study the background space--time is taken to be a
Robertson--Walker space--time with line--element
\begin{equation}\label{7.1}
ds^2=\Omega^{2}(T)\{-dT^2+{}^3g_{\alpha\beta}dx^\alpha\,dx^\beta\}\
.
\end{equation}
Here the greek indices take values $1,\,2,\,3$ and
${}^3g_{\alpha\beta}$ is the metric tensor for a three--space of
constant curvature. Comparing this to (\ref{3.4}) we see that our
background space--time also has this form if we take
${}^3g_{\alpha\beta}=(p_0^{-2}f^2,\,p_0^{-2}f^2,\,1)$ and label
the coordinates $x^1=y,\,x^2=z,\,x^3=x$. The method used in
\cite{bardeen} involves separating the time dependent and spatial
dependent parts of the perturbations. Now for us the important
coordinates are $u=(x-T(t))/\sqrt{2}$, $v=(x+T(t))/\sqrt{2}$ and
there is no natural way to carry out this separation. Thus it is
not possible to directly compare the results found here with those
of \cite{bardeen}. However there are some obvious similarities and
differences between the results and we shall briefly comment on
these now.  One point of agreement is that gravitational radiation
is described by tensor perturbations only. Specifically in our
case gravitational waves are described by perturbations in the
shear and anisotropic stress tensors. The perturbed space--time in
\cite{bardeen} is given by
\begin{equation} \label{7.2}
ds^2=-\Omega^2dT^2+g_{\alpha\beta}dx^\alpha\,dx^\beta\ ,
\end{equation}
where
\begin{equation} \label{7.3}
g_{\alpha\beta}=\Omega^2[{}^3g_{\alpha\beta}+2H^{(2)}_T(T)\,Q^{(2)}_{\alpha\beta}(x^\mu)]\
,
\end{equation}
and $Q^{(2)}_{\alpha\beta}$ is a divergenceless trace--free
tensor. This bears a strong resemblance to our perturbed
space--time described by (\ref{6.1}) where $\gamma_{ab}$ given in
(\ref{6.2}) is also divergenceless and trace-free. However it is
clear from (\ref{6.2}) that, in effect, our small metric
perturbations $\gamma_{ab}$ are expressible in the form of a
$2\times 2$ matrix whereas $(Q^{(2)}_{\alpha\beta})$ is a $3\times
3$ matrix. In addition $\gamma_{ab}$ satisfies the inhomogeneous
wave equation (\ref{6.12}) while $Q^{(2)}_{\alpha\beta}$ satisfies
the homogeneous wave equation \cite{bardeen}
\begin{equation} \label{7.4}
Q^{(2)\,\alpha\beta;\gamma}{}_{;\gamma}+k_0^{2}Q^{(2)\,\alpha\beta}=0\
,
\end{equation}
where $k_0$ is a constant.

\noindent
\section*{Acknowledgment}\noindent
I thank Professor Peter Hogan for many helpful discussions in the
course of this work and IRCSET and Enterprise Ireland for
financial support.

\appendix
\section{The Ricci Tensor Components} \setcounter{equation}{0}
In this section we give the Ricci tensor components (on the tetrad
given by Eqs. (\ref{5.8})) for the metric defined by Eqs.
(\ref{5.8}) and (\ref{5.9}). In the calculation of the Ricci
tensor components we use $\partial \Omega/\partial u = -\partial
\Omega/\partial v$ and $\partial f/\partial u = \partial
f/\partial v$ to simplify equations. Also for convenience we shall
use subscripts $y,z,u,v$ to indicate partial derivatives with
respect to these variables and a prime to denote partial
differentiation with respect to $v$. Following the Cartan method
to find the Ricci tensor components we first find the non--zero
Ricci rotation coefficients to be:
\begin{eqnarray}\label{A.1}
\Upsilon_{121} &=&
-\frac{1}{2}\,\Omega^{-1}\,f^{-1}\,(1+\alpha)\,K\,z+\Omega^{-1}\,f^{-1}\,p_{0}\,\alpha_{z}+
\frac{1}{2}\,\Omega^{-1}\,f^{-1}\,K\,\beta\,y
\nonumber \\ & &-\Omega^{-1}\,f^{-1}\,p_{0}\,\beta_{y}\ ,  \\
\Upsilon_{131}&=&
-\Omega^{-2}\,\Omega^{'}+\Omega^{-1}\,f^{-1}\,f^{'}+\Omega^{-1}\,\alpha_{u}\
,  \\
\Upsilon_{141} &=&
\Omega^{-2}\,\Omega^{'}+\Omega^{-1}\,f^{-1}\,f^{'}+\Omega^{-1}\,\alpha_{v}\
, \\
\Upsilon_{212}&=&\frac{1}{2}\,\Omega^{-1}\,f^{-1}\,K\,\beta\,z-\Omega^{-1}\,f^{-1}\,p_{0}\,\beta_{z}
-\,\frac{1}{2}\Omega^{-1}\,f^{-1}\,(1-\alpha)\,K\,y\nonumber
\\ & &-\Omega^{-1}\,f^{-1}\,p_{0}\,\alpha_{y}\
,  \\
\Upsilon_{232}&=& -\Omega^{-2}\,\Omega^{'} +
\Omega^{-1}\,f^{-1}\,f^{'}-\Omega^{-1}\,\alpha_{u}\ , \\
\Upsilon_{242}&=& \Omega^{-2}\,\Omega^{'} +
\Omega^{-1}\,f^{-1}\,f^{'}-\Omega^{-1}\,\alpha_{v}\ ,  \\
\Upsilon_{343}&=& \Omega^{-2}\,\Omega^{'}\ ,  \\
\Upsilon_{434}&=& -\Omega^{-2}\,\Omega^{'}\ ,  \\
\Upsilon_{231} &=&\Upsilon_{132} = \Omega^{-1}\,\beta_{u}\ ,\\
\Upsilon_{142}&=&\Upsilon_{241}=\Omega^{-1}\,\beta_{v}\ .
\end{eqnarray}
We note that in this calculation we have discarded any terms which
are second order or smaller in $\alpha, \beta$. Using these
coefficients we obtain the Ricci tensor components:
\begin{eqnarray}\label{A.11} R_{13} &=&
p_{0}^{3}\,f^{-1}\,\Omega^{-2}\{(p_0^{-2}\,\alpha)_{yu}+(p_{0}^{-2}\,\beta)_{zu}\}\
,  \\
R_{23} &=&
-\,p_{0}^{3}\,f^{-1}\,\Omega^{-2}\{(p_0^{-2}\,\alpha)_{zu}-(p_{0}^{-2}\,\beta)_{yu}\}\
,  \\
R_{33} &=&
4\,\Omega^{-4}\,\Omega^{'2}-2\,\Omega^{-3}\,\Omega^{''}-2\,f^{-1}\,f^{''}\,\Omega^{-2}\
,  \\
R_{44} &=&
4\,\Omega^{-4}\,\Omega^{'2}-2\,\Omega^{-3}\,\Omega^{''}-2\,f^{-1}\,f^{''}\,\Omega^{-2}\
, \\
R_{34} &=&
4\,\Omega^{-3}\,\Omega^{''}-2\,\Omega^{-4}\,\Omega^{'2}-2\,f^{-1}\,f^{''}\,\Omega^{-2}\
, \\
R_{14}&=&
p_{0}^{3}\,f^{-1}\,\Omega^{-2}\{(p_{0}^{-2}\alpha)_{yv}+(p_{0}^{-2}\beta)_{zv}\}\
, \\
R_{24}&=&
-\,p_{0}^{3}\,f^{-1}\,\Omega^{-2}\{(p_{0}^{-2}\alpha)_{zv}-(p_{0}^{-2}\beta)_{yv}\}\
, \\
R_{12} &=&
2\,\Omega^{-3}\,\Omega^{'}\,(\beta_{v}-\beta_{u})-2\,\Omega^{-2}\,\beta_{uv}-2\,f^{-1}\,f^{'}\,\Omega^{-2}(\beta_{u}+\beta_{v})\
, \\
R_{11} &=&
2\,\Omega^{-4}\,\Omega^{'2}-2\,f^{-1}\,f^{''}\,\Omega^{-2}+2\,\Omega^{-3}\,\Omega^{''}-2\Omega^{-2}f^{-2}f^{'2}+\Omega^{-2}f^{-2}K
\nonumber \\
&+&2\Omega^{-3}\Omega^{'}(\alpha_v-\alpha_u)-2\Omega^{-2}f^{-1}f^{'}(\alpha_v+\alpha_u)-2\Omega^{-2}\alpha_{uv}
\nonumber \\
&+&
\Omega^{-2}f^{-2}\{p_0^{4}(p_0^{-2}\alpha)_{yy}-p_0^{4}(p_0^{-2}\alpha)_{zz}-Kzp_0^{3}(p_0^{-2}\alpha)_z+Kyp_0^3(p_0^{-2}\alpha)_y\}
\nonumber \\
&+&
\Omega^{-2}f^{-2}\{2p_0^{4}(p_0^{-2}\beta)_{yz}+Kzp_0^{3}(p_0^{-2}\beta)_y+Kyp_0^3(p_0^{-2}\beta)_z\}\
,
\\
R_{22}
&=&2\,\Omega^{-4}\,\Omega^{'2}-2\,f^{-1}\,f^{''}\,\Omega^{-2}+2\,\Omega^{-3}\,\Omega^{''}-2\Omega^{-2}f^{-2}f^{'2}+\Omega^{-2}f^{-2}K
\nonumber \\
&-&2\Omega^{-3}\Omega^{'}(\alpha_v-\alpha_u)+2\Omega^{-2}f^{-1}f^{'}(\alpha_v+\alpha_u)+2\Omega^{-2}\alpha_{uv}
\nonumber \\
&+&
\Omega^{-2}f^{-2}\{p_0^{4}(p_0^{-2}\alpha)_{yy}-p_0^{4}(p_0^{-2}\alpha)_{zz}-Kzp_0^{3}(p_0^{-2}\alpha)_z+Kyp_0^3(p_0^{-2}\alpha)_y\}
\nonumber \\
&+&
\Omega^{-2}f^{-2}\{2p_0^{4}(p_0^{-2}\beta)_{yz}+Kzp_0^{3}(p_0^{-2}\beta)_y+Kyp_0^3(p_0^{-2}\beta)_z\}\
 \ .
\end{eqnarray}

\section{The Existence of Gauge Terms if $a_1^A$, $a_2^{A}$ are non--zero}
\setcounter{equation}{0} In this Appendix we demonstrate that
$a_1^{A}$, $a_2^{A}$ appearing in (\ref{5.6}) are pure gauge
terms. For clarity we shall consider only cases when $f^{'}\neq0$.
When $a^A_1\neq0$, $a^A_2\neq0$ the line--element (\ref{5.6}) with
$b$, $c$ given in (\ref{5.7}) can be written in the form
\begin{eqnarray} \label{B.1}
ds^2&=&2\,\Omega^2\,du\,dv+p_{0}^{-2}\,f^2\,\Omega^2\{(1+\alpha)dy+\beta\,dz+A\,du+P\,dv\}^2
\nonumber \\
&+&p_{0}^{-2}\,f^2\,\Omega^2\{\beta\,dy+(1-\alpha)dz+B\,du+Q\,dv\}^2
\ ,
\end{eqnarray}
where
\begin{eqnarray} \label{B.2}
A&=&a^1_{1}\,e^{\alpha}\,\cosh{\beta}+a^2_1\,e^{-\alpha}\,\sinh{\beta}\
, \nonumber \\
B&=&a^1_{1}\,e^{\alpha}\,\sinh{\beta}+a^2_1\,e^{-\alpha}\,\cosh{\beta}\
,
\nonumber \\
P&=&a^1_{2}\,e^{\alpha}\,\cosh{\beta}+a^2_2\,e^{-\alpha}\,\sinh{\beta}\
, \nonumber \\
Q&=&a^1_{2}\,e^{\alpha}\,\sinh{\beta}+a^2_2\,e^{-\alpha}\,\cosh{\beta}\
.
\end{eqnarray}
We find it convenient to work on the following
tetrad:
\begin{eqnarray} \label{B.3}
\theta^1&=&p^{-1}_{0}\,f\,\Omega\{(1+\alpha)dy+\beta\,dz+A\,du+P\,dv\}\
, \nonumber \\
\theta^2&=&p^{-1}_{0}\,f\,\Omega\{\beta\,dy+(1-\alpha)\,dz+B\,du+Q\,dv\}\
, \nonumber \\
\theta^3&=&\Omega\,du\ , \nonumber \\
\theta^4&=&\Omega\,dv\ .
\end{eqnarray}
As before our first step is to calculate the Ricci tensor
components. In this case they are found to be:
\begin{eqnarray}
R_{13}&=&2\,p_0^{-1}f^{'}\Omega^{-2}(P_u-A_v)-
p_{0}^{-1}f\Omega^{-3}\Omega^{'}(P_u-A_v) \nonumber \\
&+&\frac{1}{2}p_0^{-1}f\Omega^{-2}(P_{u}-A_{v})_u+p_{0}^{-3}f^{-1}\Omega^{-2}\{(p_{0}^{-2}\alpha)_{yu}+(p_{0}^{-2}\beta)_{zu}\}
\nonumber
\\
&-&\frac{1}{2}p_{0}^{3}f^{-1}\Omega^{-2}\left(B_{z}Kp_{0}^{-3}y-\frac{1}{2}BK^{2}p_{0}^{-4}y\,z-p_{0}^{-2}B_{yz}\right)
\nonumber \\
&-&\frac{1}{2}p_{0}^{3}f^{-1}\Omega^{-2}\left(p_0^{-2}A_{zz}-p_{0}^{-3}KA_z\,z+AKp_{0}^{-3}
-\frac{1}{2}AK^{+2}p_{0}^{-4}y^2\right)\ ,\\
\nonumber\\
R_{23}&=&2p_0^{-1}f^{'}\Omega^{-2}(Q_u-B_v)-
p_{0}^{-1}f\Omega^{-3}\Omega^{'}(Q_u-B_v) \nonumber \\
&+&\frac{1}{2}p_0^{-1}f\Omega^{-2}(Q_{u}-B_{v})_u-p_{0}^{-3}f^{-1}\Omega^{-2}\{(p_{0}^{-2}\alpha)_{zu}-(p_{0}^{-2}\beta)_{yu}\}
\nonumber \\
&-&\frac{1}{2}p_{0}^{3}f^{-1}\Omega^{-2}\left(A_{y}Kp_{0}^{-3}z-\frac{1}{2}AK^{2}p_{0}^{-4}y\,z-p_{0}^{-2}A_{yz}\right)
\nonumber \\
&-&\frac{1}{2}p_{0}^{3}f^{-1}\Omega^{-2}\left(p_0^{-2}B_{yy}-p_{0}^{-3}KB_{y}z+BKp_{0}^{-3}
-\frac{1}{2}BK^{2}p_{0}^{-4}z^2\right)\ , \\ \nonumber \\
R_{33}&=&4\,\Omega^{-4}\,\Omega^{'2}-2\,\Omega^{-3}\,\Omega^{''}-2\,\Omega^{-2}\,f^{-1}\,f^{''}+p_{0}^{2}\,\Omega^{-2}\{
(p_{0}^{-2}\,A)_{yu}+(p_{0}^{-2}\,B)_{zu}\}\nonumber \\
&+&2\Omega^{-2}f^{-1}f^{'}p_0^{2}\{(p_0^{-2}A)_y+(p_0^{-2}B)_z\}\ , \\ \nonumber \\
R_{44}&=&4\,\Omega^{-4}\,\Omega^{'2}-2\,\Omega^{-3}\,\Omega^{''}-2\,\Omega^{-2}\,f^{-1}\,f^{''}+p_0^{-2}\,\Omega^{-2}\{(p_0^{-2}\,P)_{yv}+
(p_{0}^{-2}\,Q)_{zv}\}\nonumber \\
&+&2\,\Omega^{-2}\,f^{-1}\,f^{'}p_0^{2}\{(p_0^{-2}Q)_z+(p_{0}^{-2}\,P)_y\}\
, \\ \nonumber \\
R_{34}&=&4\,\Omega^{-3}\,\Omega^{''}-2\,\Omega^{-4}\,\Omega^{'2}-2\,\Omega^{-2}\,f^{-1}\,f^{''}+p_{0}^{2}\,\Omega^{-3}\Omega^{'}
\{(p_{0}^{-2}A)_y+(p_{0}^{-2}B)_z\}\nonumber \\
&-&p_{0}^{2}\,\Omega^{-3}\,\Omega^{'}\{(p_{0}^{-2}P)_y+(p_{0}^{-2}Q)_z\}+\frac{1}{2}p_{0}^{2}\,\Omega^{-2}\{
(p_0^{-2}A)_{yv}+(p_0^{-2}B)_{zv}\}\nonumber \\
&+&\frac{1}{2}p_{0}^{2}\,\Omega^{-2}\{(p_0^{-2}P)_{yu}+(p_0^{-2}Q)_{zu}\}+p_{0}^{2}\,f^{-1}\,f^{'}\,\Omega^{-2}\{(p_0^{-2}A)_y+(p_0^{-2}B)_z\}
\nonumber\\
&+&p_{0}^{2}\,f^{-1}\,f^{'}\,\Omega^{-2}\{(p_0^{-2}P)_y+(p_0^{-2}Q)_z\}\
, \\ \nonumber \\
R_{14}&=&p_0^{3}f^{-1}\Omega^{-2}\{(p_0^{-2}\alpha)_{yv}+(p_0^{-2}\beta)_{zv}\}-p_0^{-1}f\Omega^{-3}\Omega^{'}(P_u-A_v)
\nonumber \\
&-&2p_0^{-1}f^{'}\Omega^{-2}(P_u-A_v)-\frac{1}{2}p_0^{-1}f\Omega^{-2}(P_u-A_v)_v\nonumber \\
&-&\frac{1}{2}p_0^{3}f^{-1}\Omega^{-2}\left\{p_0^{-2}P_{zz}-P_zKp_0^{-3}\,z+PKp_0^{-3}-\frac{1}{2}PK^{2}p_0^{-4}\,y^2\right\}
\nonumber \\
&-&\frac{1}{2}p_0^{3}f^{-1}\Omega^{-2}\left\{Q_zKp_0^{-3}\,y-p_0^{-2}Q_{yz}-\frac{1}{2}QK^2p_0^{-4}\,y\,z\right\}\
,\\ \nonumber \\
R_{24}&=&-p_0^{3}f^{-1}\Omega^{-2}\{(p_0^{-2}\alpha)_{zv}-(p_0^{-2}\beta)_{yv}\}-p_0^{-1}f\Omega^{-3}\Omega^{'}(Q_u-B_v)
\nonumber \\
&-&2p_0^{-1}f^{'}\Omega^{-2}(Q_u-B_v)-\frac{1}{2}p_0^{-1}f\Omega^{-2}(Q_u-B_v)_v\nonumber \\
&-&\frac{1}{2}p_0^{3}f^{-1}\Omega^{-2}\left\{p_0^{-2}Q_{yy}-Q_yKp_0^{-3}\,y+QKp_0^{-3}-\frac{1}{2}QK^{2}p_0^{-4}\,z^2\right\}
\nonumber \\
&-&\frac{1}{2}p_0^{3}f^{-1}\Omega^{-2}\left\{P_yKp_0^{-3}\,z-p_0^{-2}P_{yz}-\frac{1}{2}PK^2p_0^{-4}\,y\,z\right\}\
,\\ \nonumber \\
R_{11}&=&\Omega^{-2}f^{-2}\{p_0^{4}(p_0^{-2}\alpha)_{zz}-p_0^{4}(p_0^{-2}\alpha)_{yy}+Kzp_0^{3}(p_0^{-2}\alpha)_z
-Kyp_0^3(p_0^{-2}\alpha)_y\}\nonumber\\
&+&\Omega^{-2}f^{-2}\{2p_0^{4}(p_0^{-2}\beta)_{yz}+Kzp_0^{3}(p_0^{-2}\beta)_y+Kyp_0^3(p_0^{-2}\beta)_z\}+\Omega^{-2}f^{-2}K
\nonumber \\
&+&(p_0^{2}\Omega^{-3}\Omega^{'}+p_0^{2}f^{-1}f^{'}\Omega^{-2})\{3p_0^{-2}A_y-2AKp_0^{-3}\,y-2BKp_0^{-3}\,z+p_0^{-2}B_z\}\nonumber\\
&-&(p_0^{2}\Omega^{-3}\Omega^{'}-p_0^{2}f^{-1}f^{'}\Omega^{-2})\{3p_0^{-2}P_y-2PKp_0^{-3}\,y-2QKp_0^{-3}\,z+p_0^{-2}Q_z\}\nonumber
\\
&-&\frac{1}{2}p_0^{-1}\Omega^{-2}K\,z(B_v+Q_u)+\Omega^{-2}\left(A_{yv}-\frac{1}{2}A_vKp_0^{-1}\,y+P_{uy}-\frac{1}{2}P_uKp_0^{-1}\,y\right)
\nonumber \\
&-&2\Omega^{-2}\alpha_{uv}-2\Omega^{-2}f^{-2}f^{'2}+2\Omega^{-4}\Omega^{'2}-2\Omega^{-2}f^{-1}f^{''}+2\Omega^{-3}\Omega^{''}
\nonumber \\&-&2\Omega^{-3}\Omega^{'}(\alpha_u-\alpha_v)
-2\Omega^{-2}f^{-1}f^{'}(\alpha_v+\alpha_u)\ ,\\
\nonumber \\
R_{22}&=&\Omega^{-2}f^{-2}\{p_0^{4}(p_0^{-2}\alpha)_{zz}-p_0^{4}(p_0^{-2}\alpha)_{yy}+Kzp_0^{3}(p_0^{-2}\alpha)_z
-Kyp_0^3(p_0^{-2}\alpha)_y\}\nonumber\\
&+&\Omega^{-2}f^{-2}\{2p_0^{4}(p_0^{-2}\beta)_{yz}+Kyp_0^{3}(p_0^{-2}\beta)_z+Kzp_0^3(p_0^{-2}\beta)_y\}+\Omega^{-2}f^{-2}K
\nonumber \\
&+&(p_0^{2}\Omega^{-3}\Omega^{'}+p_0^{2}f^{-1}f^{'}\Omega^{-2})\{3p_0^{-2}B_z-2BKp_0^{-3}\,z-2AKp_0^{-3}\,y+p_0^{-2}A_y\}\nonumber\\
&-&(p_0^{2}\Omega^{-3}\Omega^{'}-p_0^{2}f^{-1}f^{'}\Omega^{-2})\{3p_0^{-2}Q_z-2QKp_0^{-3}\,z-2PKp_0^{-3}\,y+p_0^{-2}P_y\}\nonumber
\\
&-&\frac{1}{2}p_0^{-1}\Omega^{-2}K\,y(A_v+P_u)+\Omega^{-2}\left(B_{zv}-\frac{1}{2}B_vKp_0^{-1}\,z+Q_{uz}-\frac{1}{2}Q_uKp_0^{-1}\,z\right)
\nonumber \\
&+&2\Omega^{-2}\alpha_{uv}-2\Omega^{-2}f^{-2}f^{'2}+2\Omega^{-4}\Omega^{'2}-2\Omega^{-2}f^{-1}f^{''}+2\Omega^{-3}\Omega^{''}
\nonumber \\&+&2\Omega^{-3}\Omega^{'}(\alpha_u-\alpha_v)
+2\Omega^{-2}f^{-1}f^{'}(\alpha_v+\alpha_u)\ ,\\
\nonumber \\
R_{12}&=&\Omega^{-3}\Omega^{'}(A_z+B_y)-\Omega^{-3}\Omega^{'}(P_z+Q_y)+2\Omega^{-3}\Omega^{'}(\beta_v-\beta_u)-2\Omega^{-2}\beta_{uv}
\nonumber \\
&-&2\Omega^{-2}f^{-1}f^{'}(\beta_u+\beta_v)+\Omega^{-2}f^{-1}f^{'}(A_z+B_y)+\Omega^{-2}f^{-1}f^{'}(Q_y+P_z)\nonumber
\\
&+&\frac{1}{2}\Omega^{-2}(A_{zv}+B_{yv})+\frac{1}{2}\Omega^{-2}(P_{zu}+Q_{yu})\
.
\end{eqnarray}
Here the subscripts $y,z,u,v$ indicate partial differentiation
with respect to these variables, differentiation with respect to
$v$ is denoted by a prime and $K$ is the constant introduced in
Eq. (\ref{3.4}). Using the above Ricci tensor components and Eqs.
(\ref{3.2}), (\ref{3.3}), (\ref{5.11}) it is easily checked that
the first of the field equations (\ref{2.11}) is satisfied
provided we choose $A,B,P,Q$ to satisfy the Cauchy--Riemann
equations
\begin{equation}\label{B.14}
(p_0^{-2}\,A)_z=(p_0^{-2}B)_y\ , \qquad
(p_0^{-2}\,A)_y=-(p_0^{-2}\,B)_z\ ,
\end{equation}
\begin{equation} \label{B.15}
(p_0^{-2}\,P)_z=(p_0^{-2}Q)_y\ , \qquad
(p_0^{-2}\,P)_y=-(p_0^{-2}\,Q)_z\ .
\end{equation}
Next with $A,B,P,Q$ satisfying these equations we find from the
remaining two equations in (\ref{2.11}) that the conditions
\begin{eqnarray}\label{B.16}
p_0^{-4}AK &=& (p_0^{-2}\alpha)_{yu}+(p_0^{-2}\beta)_{zu}\ , \\
p_0^{-4}P\,K &=&(p_0^{-2}\alpha)_{yv}+(p_0^{-2}\beta)_{zv}\
, \\
p_0^{-4}BK &=& -(p_0^{-2}\alpha)_{zu}+(p_0^{-2}\beta)_{yu}\ , \\
p_0^{-4}Q\,K &=& -(p_0^{-2}\alpha)_{zv}+(p_0^{-2}\beta)_{yv}\ , \\
P_u&=&A_v\ , \qquad Q_u=B_v\ ,
\end{eqnarray}
are sufficient to have $q_a\equiv0$ and $\pi_{ab}=0$ except for
\begin{eqnarray} \label{B.21}
\pi_{11}&=&
(\frac{1}{\sqrt{2}}R^{-2}\dot{R}p_0^2+R^{-2}f^{-1}f^{'}p_0^{2})\{3p_0^{-2}A_y-2AKp_0^{-3}\,y-2BKp_0^{-3}\,z+p_0^{-2}\,B_z\}
\nonumber \\
&-&(\frac{1}{\sqrt{2}}R^{-2}\dot{R}p_0^2-R^{-2}f^{-1}f^{'}p_0^{2})\{3p_0^{-2}P_y-2PKp_0^{-3}\,y-2QKp_0^{-3}\,z+p_0^{-2}\,Q_z\}
\nonumber \\
&-&\frac{1}{2}R^{-2}\,Kp_0^{-1}\,z(B_v+Q_u)+R^{-2}\left(A_{yv}-\frac{1}{2}A_v\,K\,p_0^{-1}\,y+P_{uy}-\frac{1}{2}P_u\,Kp_0^{-1}\,y\right)
\nonumber \\
&+&R^{-2}f^2\{p_0^{4}(p_0^{-2}\alpha)_{zz}-p_0^{4}(p_0^{-2}\alpha)_{yy}+Kzp_0^{3}(p_0^{-2}\alpha)_z
-Kyp_0^3(p_0^{-2}\alpha)_y\}\nonumber
\\
&+&R^{-2}f^2\{2p_0^{4}(p_0^{-2}\beta)_{yz}+Kzp_0^{3}(p_0^{-2}\beta)_y+Kyp_0^3(p_0^{-2}\beta)_z\}
-2R^{-2}\,\alpha_{vu}\nonumber
\\
&+&\frac{2}{\sqrt{2}}R^{-2}\dot{R}(\alpha_v-\alpha_u)-2R^{-2}f^{-1}f^{'}(\alpha_v+\alpha_u)\
, \\ \nonumber \\ \nonumber \\
\pi_{22}&=&(\frac{1}{\sqrt{2}}R^{-2}\dot{R}p_0^2+R^{-2}f^{-1}f^{'}p_0^{2})\{3p_0^{-2}B_z-2BKp_0^{-3}\,z-2AKp_0^{-3}\,y+p_0^{-2}\,A_y\}
\nonumber \\
&-&(\frac{1}{\sqrt{2}}R^{-2}\dot{R}p_0^2-R^{-2}f^{-1}f^{'}p_0^{2})\{3p_0^{-2}Q_z-2QKp_0^{-3}\,z-2PKp_0^{-3}\,y+p_0^{-2}\,P_y\}
\nonumber \\
&-&\frac{1}{2}R^{-2}\,Kp_0^{-1}\,z(A_v+P_u)+R^{-2}\left(B_{zv}-\frac{1}{2}B_v\,K\,p_0^{-1}\,z+P_{uz}-\frac{1}{2}Q_u\,Kp_0^{-1}\,z\right)
\nonumber \\
&+&R^{-2}f^2\{p_0^{4}(p_0^{-2}\alpha)_{zz}-p_0^{4}(p_0^{-2}\alpha)_{yy}+Kzp_0^{3}(p_0^{-2}\alpha)_z
-Kyp_0^3(p_0^{-2}\alpha)_y\}\nonumber
\\
&+&R^{-2}f^2\{2p_0^{4}(p_0^{-2}\beta)_{yz}+Kzp_0^{3}(p_0^{-2}\beta)_y+Kyp_0^3(p_0^{-2}\beta)_z\}+
2R^{-2}\,\alpha_{vu}\nonumber
\\
&+&\frac{2}{\sqrt{2}}R^{-2}\dot{R}(\alpha_u-\alpha_v)+2R^{-2}f^{-1}f^{'}(\alpha_v+\alpha_u)\
, \\ \nonumber \\
\pi_{12}&=&\frac{1}{\sqrt{2}}R^{-2}\,\dot{R}(A_z+B_y-P_z-Q_y)+R^{-2}\,f^{-1}\,f^{'}(A_z+B_y+P_z+Q_y)
\nonumber \\
&+&\frac{1}{2}R^{-2}(A_{zv}+B_{yv}+P_{zu}+Q_{yu})+\sqrt{2}R^{-2}\dot{R}(\beta_v-\beta_u)-2R^{-2}\,\beta_{uv}\nonumber
\\
&-&2R^{-2}f^{-1}f^{'}(\beta_u+\beta_v)\ .
\end{eqnarray}
Following the procedure described in Section 5 we now construct
$\bar{\pi}=(\pi_{11}-\pi_{22}-2i\pi_{12})/2$. Using the conditions
above to cancel terms we arrive at
\begin{eqnarray} \label{B.24}
\bar{\pi}&=&\sqrt{2}\,R^{-2}\,\dot{R}\{\alpha_v-\alpha_u-i(\beta_v-\beta_u)\}-2R^{-2}f^{-1}f^{'}\{\alpha_v+\alpha_u-i(\beta_v+\beta_u)\}
\nonumber \\
&+&\frac{1}{\sqrt{2}}R^{-2}\dot{R}\{A_y-B_z+Q_z-P_y-i(A_z+B_y-P_z-Q_y)\}-2R^{-2}(\alpha_{uv}-i\beta_{uv})\nonumber
\\
&+&R^{-2}f^{-1}f^{'}\{A_y-B_z+P_y-Q_z-i(A_z+B_y+P_z+Q_y)\}\nonumber
\\
&+&\frac{1}{2}\,R^{-2}\{A_{yv}-B_{zv}+P_{yu}-Q_{zu}-i(A_{zv}+B_{yv}+P_{zu}+Q_{yu})\}\
.
\end{eqnarray}
We want to write $\bar{\pi}$ in the form given in Eq. (\ref{5.26})
for some analytic function ${\cal{G}}$. Before we try to do this
we note that on account of the conditions $P_u=A_v$, $Q_u=B_v$ we
can write
\begin{eqnarray} \label{B.25}
A&=&F_u\ , \qquad P=F_v\ , \\
B&=&G_u\ , \qquad Q=G_v\ ,
\end{eqnarray}
for some functions $F(y,z,u,v)$, $G(y,z,u,v)$ which satisfy the
Cauchy--Riemann equations
\begin{equation} \label{B.27}
(p_0^{-2}F)_y=-(p_0^{-2}G)_z\ , \qquad
(p_0^{-2}F)_z=(p_0^{-2}G)_y\ .
\end{equation}
Substituting these into Eq. (\ref{B.24}) gives
\begin{eqnarray} \label{B.28}
\bar{\pi}&=&\frac{1}{\sqrt{2}}R^{-2}\dot{R}\{F_{uy}-G_{uz}+G_{vz}-F_{vy}-i(F_{uz}+G_{uy}-F_{vz}-G_{vy})\}\nonumber \\
&+&R^{-2}f^{-1}f^{'}\{F_{uy}-G_{uz}+F_{vy}-G_{vz}-i(F_{uz}+G_{uy}+F_{vz}+G_{vy})\}\nonumber
\\
&+&R^{-2}\{F_{uyv}-G_{uvz}-i(F_{uvz}+G_{uvz})\}+\sqrt{2}R^{-2}\dot{R}\{\alpha_v-\alpha_u-i(\beta_v-\beta_u)\}\nonumber
\\
&-&2R^{-2}f^{-1}f^{'}\{\alpha_v+\alpha_u-i(\beta_v+\beta_u)\}-2R^{-2}(\alpha_{uv}-i\beta_{uv})\
.
\end{eqnarray}
In order to write $\bar{\pi}$ in the required form we choose
\begin{eqnarray}\label{B.29}
{\cal{G}}&=&\frac{1}{2\sqrt{2}}p_0^{-2}\,f\{G_{uz}-F_{uy}-G_{vz}+F_{vy}+i(F_{uz}+G_{uy}-F_{vz}-G_{vy})\}\nonumber
\\
&-&\frac{1}{\sqrt{2}}p_0^{-2}\,f\{\alpha_{v}-\alpha_{u}+i(\beta_{u}-\beta_{v})\}\
.
\end{eqnarray}
Noting that $D=\sqrt{2}\,\partial_v$ we find that this ${\cal{G}}$
does satisfy Eq. (\ref{5.26}) with $\bar{\pi}$ given by
(\ref{B.28}) and $f^{'}\neq0$ provided $\alpha$, $\beta$ take the
following form:
\begin{equation}\label{30}
\alpha = \frac{1}{2}(F_y-G_z)+q(y,z,u)\ , \qquad
\beta=\frac{1}{2}(F_z+G_y)+r(y,z,u)\ .
\end{equation}
Here $q$, $r$ satisfy the Cauchy--Riemann equations
\begin{equation} \label{B.31}
(p_0^{-2}\,q)_y=-(p_0^{-2}\,r)_z\ , \qquad
(p_0^{-2}\,q)_z=(p_0^{-2}\,r)_y\ .
\end{equation}
We remark that this is the first time we have made use of the fact
that $f^{'}\neq0$. If $f^{'}=0$ then $\alpha$, $\beta$ have a
different form to that given in the last equation. Thus we
emphasise that the analysis which follows does not apply if
$f^{'}=0$. When $\alpha$, $\beta$ are given by these equations it
is straightforward to check using the various Cauchy--Riemann
equations that $\pi_{ab}$ is trace--free and the conditions
(B.16)--(B.19) are identically satisfied. Substituting the above
expressions for $\alpha$, $\beta$ into Eq. (\ref{B.29}) yields
\begin{equation}\label{B.32}
{\cal{G}}=\frac{1}{\sqrt{2}}p_0^{-2}\,f(q_u+ir_u)\ .
\end{equation}
With $k$ given by the first of Eqs. (\ref{5.22}) it is trivial to
show that ${\cal{G}}$ satisfies the wave equation (\ref{4.25}).
Now $A,B,P,Q$ do not appear on the right--hand side of
(\ref{B.32}) and hence $a_1^A$, $a_2^A$ do not contribute to
${\cal{G}}$. Thus since the perturbed shear and anisotropic stress
can both be written in terms of ${\cal{G}}$ we conclude that
$a_1^A$, $a_2^A$ are pure gauge terms.

\end{document}